\documentclass[floats,floatfix,showpacs,amssymb,prd,twocolumn,superscriptaddress,nofootinbib,nolongbibliography,reprint,preprintnumbers]{revtex4-2}

\usepackage{amssymb,amsmath,verbatim,mathtools,needspace,enumitem,etoolbox,graphicx,microtype,afterpage,bigints,tabularx,nicefrac,bm}
\usepackage[dvipsnames, usenames]{xcolor}
\definecolor{linkcolor}{rgb}{0.0,0.3,0.5}
\usepackage[unicode, colorlinks=true, linkcolor=linkcolor, citecolor=linkcolor, filecolor=linkcolor,urlcolor=linkcolor, pdfusetitle]{hyperref}
\usepackage{booktabs}
\usepackage[all]{hypcap}
\usepackage[T1]{fontenc}
\usepackage[utf8]{inputenc}
\usepackage{orcidlink}
\usepackage{mathrsfs}
\usepackage{siunitx}
\usepackage{soul}
\DeclareSIUnit \parsec {pc}
\DeclareSIUnit \arcsecondfull {arcsec}
\DeclareSIUnit \year{yr}
\DeclareSIUnit \day{day}
\DeclareSIUnit \hour{hr}
\DeclareSIUnit \radiant{rad}
\DeclareSIUnit \degfull{deg}
\DeclareSIUnit \erg {erg}
\DeclareSIUnit \Lsun {L_\odot}
\DeclareSIUnit \Msun {M_\odot}
\DeclareSIUnit \AstroUnit {au}

\newcommand{\ssim}{\mathchar"5218\relax\,}

\newcommand{\dd}{{\rm d}}
\newcommand{\pdv}[2]{\frac{\partial #1}{\partial #2}}
\newcommand{\vdot}{\boldsymbol{\cdot}}
\newcommand{\order}[1]{{\cal O}\left(#1\right)}
\renewcommand{\Re}{{\rm Re}}

\makeatletter
\let\oldtheequation\theequation
\renewcommand\tagform@[1]{\maketag@@@{\ignorespaces#1\unskip\@@italiccorr}}
\renewcommand\theequation{(\oldtheequation)}
\makeatother

\newcommand{\milan}{\affiliation{Dipartimento di Fisica ``G. Occhialini'', Universit\'a degli Studi di Milano-Bicocca, Piazza della Scienza 3, 20126 Milano, Italy}}
\newcommand{\infn}{\affiliation{INFN, Sezione di Milano-Bicocca, Piazza della Scienza 3, 20126 Milano, Italy}}
\newcommand{\geneva}{\affiliation{ D\'epartement de Physique Th\'eorique, Universit\'e de Gen\`eve, 24 quai Ernest Ansermet, 1211 Gen\`eve 4, Switzerland}}

\newcommand{\gwsc}{\affiliation{Gravitational Wave Science Center (GWSC), Universit\'e de Gen\`eve, CH-1211 Geneva, Switzerland}}

\newcommand{\potsdamUn}{\affiliation{Institut f\"{u}r Physik und Astronomie, Universit\"{a}t Potsdam, Haus 28, Karl-Liebknecht-Str. 24/25, 14476, Potsdam, Germany}}
\newcommand{\potsdamPl}{\affiliation{Max Planck Institute for Gravitational Physics (Albert Einstein Institute), D-14476 Potsdam, Germany}}
\newcommand{\caen}{\affiliation{Normandie Univ, ENSICAEN, UNICAEN, CNRS/IN2P3, LPC Caen, 14000 Caen, France}}
\newcommand{\brussels}{\affiliation{Institut d’Astronomie et d’Astrophysique, Universit\'e Libre de Bruxelles, CP 226, B-1050 Brussels, Belgium}}
\newcommand{\parisCit}{\affiliation{Laboratoire Univers et Th\'eories, CNRS, Observatoire de Paris, Universit\'e PSL, Universit\'e Paris Cit\'e, 5 place Jules Janssen, 92195 Meudon, France}}
\newcommand{\nikhef}{\affiliation{Nikhef – National Institute for Subatomic Physics, Science Park 105, 1098 XG Amsterdam, The Netherlands}}
\newcommand{\utrecht}{\affiliation{Institute for Gravitational and Subatomic Physics (GRASP), Utrecht University, Princetonplein 1, 3584 CC Utrecht, The Netherlands}}

\begin{document}

\title{Nuclear physics constraints from binary neutron star mergers in the \texorpdfstring{\\}{}Einstein Telescope era}

\author{Francesco Iacovelli\orcidlink{0000-0002-4875-5862}}

\geneva
\gwsc

\author{Michele Mancarella\orcidlink{0000-0002-0675-508X}}

\milan \infn

\author{Chiranjib Mondal\orcidlink{0000-0002-9238-6144}}

\caen
\brussels

\author{Anna Puecher\orcidlink{0000-0003-1357-4348}}

\nikhef \utrecht

\author{Tim Dietrich\orcidlink{0000-0003-2374-307X}}

\potsdamUn \potsdamPl

\author{Francesca Gulminelli\orcidlink{0000-0003-4354-2849}}

\caen

\author{Michele Maggiore\orcidlink{0000-0001-7348-047X}}

\geneva
\gwsc

\author{Micaela Oertel\orcidlink{0000-0002-1884-8654}}

\parisCit

\date{\today}

\begin{abstract}

The next generation of ground--based gravitational--wave detectors,  Einstein Telescope (ET) and  Cosmic Explorer (CE), present a unique opportunity to put constraints on dense matter, among many other groundbreaking scientific goals. In a recent study \cite{Branchesi:2023mws} the science case of ET was further strengthened, studying in particular the performances of different detector designs. In this paper we present a more detailed study of the nuclear physics section of that work. In particular, focusing on two different detector configurations (the  single--site triangular--shaped design and a design consisting of two widely separated ``L--shaped'' interferometers), we study the detection prospects of binary neutron star (BNS) mergers, and how they can reshape our understanding of the underlying equation of state (EoS) of dense matter.
We employ several state--of--the--art EoS models and state--of--the--art synthetic BNS merger catalogs, and we make use of the Fisher information formalism (FIM) to quantify  statistical errors on the astrophysical parameters describing individual BNS events. To check the reliability of the FIM method, we further perform a full parameter estimation for a few simulated events. Based on the uncertainties on the  tidal deformabilities associated to these events, we outline a mechanism to extract the underlying injected EoS using a recently developed meta--modelling approach within a Bayesian framework. 
Our results suggest that with $\gtrsim 500$ events with signal--to--noise ratio greater than $12$, we will be able to pin down very precisely the underlying EoS governing the neutron star matter. 

\end{abstract}

\preprint{ET-0276A-23}

\maketitle

{
  \hypersetup{linkcolor=black}
  \tableofcontents
}
\hypersetup{linkcolor=linkcolor}

\section{Introduction}

After the first detection of gravitational waves (GWs) from a binary black hole (BBH) coalescence~\cite{LIGOScientific:2016aoc} and the subsequent 
remarkable discoveries of the LIGO--Virgo--KAGRA (LVK) collaboration, GWs have become a new tool for exploring the Universe, and have already provided important results in astrophysics,  fundamental physics and cosmology (see \textit{e.g.} \cite{TheLIGOScientific:2017qsa,Monitor:2017mdv,LIGOScientific:2017ync,LIGOScientific:2020ibl,LIGOScientific:2021djp,LIGOScientific:2021psn,LIGOScientific:2021sio,LIGOScientific:2021aug}). 
Current facilities, however, are expected to  reach, possibly on a timescale ${\cal O}(10)$~yr,  the sensitivity limits allowed by their infrastructures and, to fully exploit the potential of GWs as a new tool for exploring the Universe, the GW community has developed the notion of third--generation (3G) ground--based detectors, in particular  Einstein Telescope (ET)~\cite{Hild:2008ng,Punturo:2010zz,Hild:2010id} 
in Europe and   Cosmic Explorer (CE)~\cite{Reitze:2019iox,Evans:2021gyd} in the US.
The science case for ET  has been  summarized in \cite{Maggiore:2019uih}, while a recent comprehensive study  of the ET capabilities for coalescing binaries can be found in \cite{Iacovelli:2022bbs} and, for multi--messenger observations, in 
\cite{Ronchini:2022gwk}. 

In its original design~\cite{Punturo:2010zz,Hild:2010id}, ET should be located ${\cal O}(200-300)$ meters underground in order to reduce the seismic noise, and features a triangular geometry consisting of three nested interferometric detectors; each detector  consists of two different instruments, a low--frequency interferometer working at cryogenic temperatures and a high--frequency interferometer working at room temperature, a configuration referred to as a ``xylophone''. However, the ET collaboration is currently performing a study to compare the original triangular design, with three nested detectors in the same infrastructure, with a more traditional geometry consisting of two widely spaced L--shaped detectors (that we will refer to as the ``2L'' configuration) while maintaining all other innovating concepts of ET, such as underground locations, cryogenics, and the xylophone configuration. This study involves many aspects, from  a comparison of the scientific return  of  the triangle versus 2L configurations, to a comparison of their costs, as well as  of possible  different financial architectures. A detailed comparison between configurations, from the purely scientific point of view, has been recently presented in Ref.~\cite{Branchesi:2023mws}. This work was realized in the context of the activities of the Observational Science Board (OSB) of ET,\footnote{The OSB is in charge of developing the science case and the technical tools relevant for ET. See \url{https://www.et-gw.eu/index.php/the-et-collaboration/observational-science-board} for a presentation of the structure and activities of the OSB and a repository of papers relevant for ET, produced in the context of the OSB activities.} and involved the study of a large number of aspects of the ET Science Case. 

In the present paper, we elaborate a part of the study of the nuclear physics section performed in Ref. \cite{Branchesi:2023mws}. In particular, we will focus on GWs from binary neutron star (BNS) systems since they are potentially among the best natural laboratories to test the behaviour of matter at super--nuclear densities. A GW detector can provide a measurement of the so--called tidal deformabilities of two coalescing objects, \textit{i.e.}, how they deform under the influence of an external tidal field (see \textit{e.g.} \cite{Baiotti:2016qnr} for a review of the physics we can extract from BNS observations). This is tightly related to the internal structure, since neutron stars deform under the action of gravity in a way that depends on the physics in their interior (\textit{e.g.} the  particles species  present, densities associated to the core of NSs, or possible phase transitions). 
Given an equation of state (EoS), the global structure of the NS \textit{i.e.}, the relations between mass and radius or tidal deformability is fully fixed, in the context of general relativity. Information on the underlying EoS can then be obtained through measurements of tidal deformabilities for a sequence of masses.
A measurement of the tidal deformation, that mostly affects the later stages of the coalescence (due to an accelerated inspiral and a merger at a lower frequency) and the post--merger phase (through a modification of the remnant stability and mass--ejection processes), is however difficult.
Furthermore, one can better extract  information on the so--called chirp mass and on a combination of the tidal deformabilities of the two component neutron stars  of a BNS system, while accurate information on the individual masses and tidal deformabilities is considerably more difficult to obtain. For this reason, the sensitivity improvement of 3G detectors is expected to play a crucial role \cite{Bauswein:2018bma,Carson:2019xxz,Pacilio:2021jmq,Smith:2021bqc,Gupta:2022qgg,Breschi:2022ens,Puecher:2022oiz,Huxford:2023qne}. 

We start here with performing a population study of the prospects of observing BNS mergers with  ET in two different designs for different EoSs, to assess how the choice of the underlying EoS impacts the overall number of detections as well as the statistical uncertainty in the reconstruction of masses and tidal parameters, obtained through a Fisher matrix analysis \cite{Cutler:1994ys,Vallisneri:2007ev,Rodriguez:2013mla}. To assess the adherence of the Fisher matrix approximation to a full Bayesian parameter estimation (PE), we also carry out a dedicated analysis on multiple injections, considering different waveform approximants, thus with different treatments of the tidal contributions. 

Once the simulated ``observations'' within the Fisher formalism are in place, we  focus on extracting the information on the properties of NS and the underlying EoS. We employed the so--called meta--modelling approach developed recently, based on a expansion in density of the energy per baryon~\cite{Margueron:2017eqc,Margueron:2017lup}. Since nuclear matter parameters (NMP) serve as the model parameters within this formalism, it can retain certain features of a more microscopic understanding of matter through underlying correlations imparted by experimental data. Yet, they are extremely cheap in terms of computation time which provides the advantages of agnostic approaches \textit{e.g.}, piecewise polytropes \cite{Read:2008iy}, spectral parametrization \cite{Lindblom:2010bb,Lindblom:2012zi,Lindblom:2013kra}, or non--parametric Gaussian process--based sampling \cite{Landry:2018prl,Essick:2019ldf,Landry:2020vaw}, which are generally employed in analyzing astrophysical data. Furthermore, the meta--modelling can provide the information on the composition, where crust and core have a unified description \cite{Carreau:2019zdy}, although the core is limited to assuming a purely nucleonic composition. Various questions concerning the data from low energy nuclear physics and nuclear astrophysics were addressed recently using the aforementioned technique \cite{Thi:2021jhz,Thi:2021hai, Mondal:2021vzt, Mondal:2022cva, Mondal:2023gbf}.

In the present calculation, we aim to reverse--engineer the implicit EoS used in the Fisher formalism, employing the previously explained meta--modelling. Since, in reality, we would not know the EoS realized in Nature, we chose a few  EoSs from the \textsc{CompOSE} database \cite{Typel:2013rza} that support NSs with at least 2 solar masses \cite{Antoniadis:2013pzd, Fonseca:2021wxt}, and we show the  corresponding posteriors for different NS properties, along with those for the nuclear matter parameters (NMPs), within a Bayesian formalism. We put particular attention to how the results change with different number of detections, and we compare the outcomes with different ET designs.

The paper is structured as follows: in \autoref{sec:setup_analysis} we describe our choices for the detectors configurations, the different EoSs adopted for neutron stars, the EoS modelling we will employ, the BNS population and the waveforms used to simulate the GW signals. In \autoref{sec:FIM_framework} we perform the comparison among Fisher matrix results and full Bayesian parameter estimation on selected injections. We then present in \autoref{subssec:pop_results} our results for the number of detections and accuracy in the reconstruction of tidal parameters for the chosen EoS. In \autoref{subssec:NS_properties_results} and \ref{subsec:nuc_phys_params_results} we report our results for the reconstruction of the NS properties and nuclear--physics parameters, respectively. Finally, we summarize our findings in \autoref{sec:conclusion}.

\section{Setup of the analysis}\label{sec:setup_analysis}

We here give a summary of the choices adopted in this work for the detector design, as well as for the equations of state of neutron star matter, for the nuclear meta--model, and for the population of merging BNS systems. We conclude the section with a brief overview of the waveform models used in the analysis.

\subsection{ET detector designs}
In Ref.~\cite{Branchesi:2023mws} various different designs for ET have been considered. In this work, we extend the analyses carried out in Sect.\ 6.2 of \cite{Branchesi:2023mws} focusing for convenience only on two representative configurations for ET:
\begin{itemize}
    \item[--] a triangular detector, composed of three nested interferometers with \SI{60}{\degree} opening angle and \SI{10}{\kilo\meter} long arms. This configuration will be denoted as $\Delta$ in the following;
    \item[--] two well--separated L--shaped interferometers, with \SI{15}{\kilo\meter} long arms and a relative orientation of \SI{45}{\degree}. This configuration will be denoted as 2L in the following.
\end{itemize}
In both cases, we consider the interferometers to feature a xylophone design, \textit{i.e.}, to be actually constituted of two instruments, one with optimized sensitivity at high frequencies and one at low frequencies, with the latter operating at cryogenic temperature.\footnote{The sensitivity curves are publicly available at \url{ https://apps.et-gw.eu/tds/ql/?c=16492}.} When considering the triangular design, we locate the detector in the Sos Enattos site in Sardinia, Italy, while in the 2L case, we locate again one detector in Sardinia and the other in the Meuse--Rhine Euroregion, across Belgium, Germany, and the Netherlands.\footnote{For the Sos Enattos site we choose as an example \{lat=\ang{40;31}, long=\ang{9;25}\}, while for the 
Meuse--Rhine site \{lat=\ang{50;43;23}, long=\ang{5;55;14}\}.}  

\begin{figure}[t]
    \centering
    \includegraphics[width=.45\textwidth]{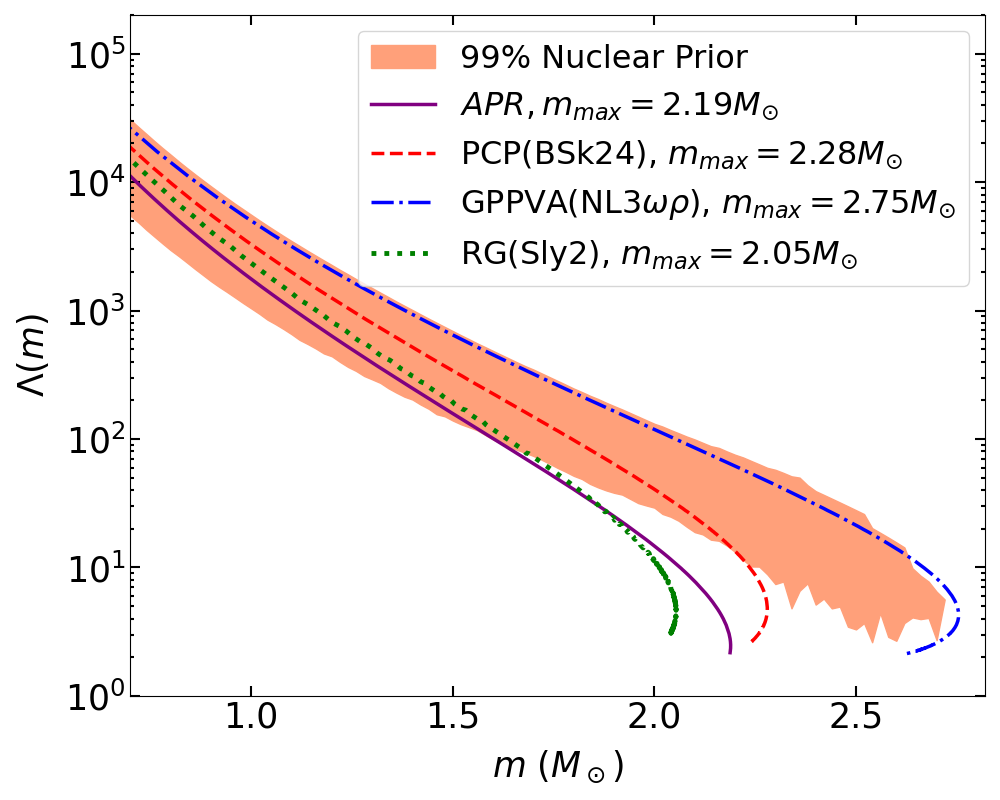}
    \caption{$\Lambda(m)$ relation for the EoS considered in the present work. We also report in the legend the maximum mass allowed by each EoS.
    }
    \label{fig:EoS_lamofm_plot}
\end{figure}

\subsection{Neutron star equations of state}
To check our capability to recover the assumed nuclear EoS and neutron
star properties, we use different representative EoS models for
obtaining the relation between tidal deformability and gravitational mass, $\Lambda(m)$, used for injection. One model,
APR~\cite{Akmal:1998cf}, is based on a fit to variational calculations
of homogeneous nuclear matter starting from realistic nucleon--nucleon
interactions for the core. The inner crust is described by the model
of Douchin and Haensel~\cite{Douchin:2001sv} and, for the outer crust,
the work by Baym, Pethick and Sutherland~\cite{Baym:1971pw} has been
used. The three other models -- RG(SLy2)~\cite{Gulminelli:2015csa},
GPPVA(NL3$\omega\rho$)~\cite{Grill:2014aea,Pais:2016xiu} and
PCP(BSk24)~\cite{Pearson:2018tkr} -- are based on nuclear density
functionals and are unified, \textit{i.e.}, NS crust and core have been
obtained from the same underlying nuclear interaction with a
consistent crust--core transition which avoids in particular
uncertainties in the predicted NS radius and tidal
deformability~\cite{Fortin:2016hny,Suleiman:2021hre}. The models have
been chosen to cover a certain range of maximum masses between \SI{2.05}{\Msun} and \SI{2.75}{\Msun} and nuclear matter and neutron star
properties, albeit remaining compatible with present
constraints. Hence, two of the models (APR and RG(SLY2)) predict NS tidal deformabilities at the
lower end of the nuclear prior, see \autoref{sec:metamodel}, one at
the upper end (GPPVA(NL3$\omega\rho$)), whereas the predictions from
the PCP(BSk24) lies well inside the region covered by the nuclear
prior. All EoS models are available in tabulated form in the
\textsc{CompOSE} database~\cite{Typel:2013rza,CompOSECoreTeam:2022ddl}.
\subsection{Nuclear meta--model for the analysis}
\label{sec:metamodel}
Our analysis to extract information on nuclear properties from the
simulated distribution in measured neutron star masses and tidal
deformabilities will be based on a nuclear meta--model which allows, in
particular, to incorporate as a prior our knowledge on the neutron
star EoS from nuclear physics and other astrophysical
sources such as pulsar mass measurements. Thus, we first generate a
prior distribution of EoSs by Monte--Carlo sampling of a large parameter
set of 17 independent, uniformly distributed empirical
parameters. These parameters characterize the density dependence of
the energy in symmetric matter (\textit{i.e.}\ equal number of protons and
neutrons) and of the symmetry energy (\textit{i.e.}\ the variation of binding
energy as a function of the neutron--to--proton ratio). 
To name a few, the most important ones corresponding to the symmetric matter are $E_{sat}$, the energy at saturation, and $K_{sat}$, the incompressibility connected to the second derivative of energy with respect to density, both evaluated at the saturation density $n_{sat}$. Similarly, the ones of foremost importance for the density dependence of the symmetry energy are the $E_{sym},L_{sym}$ and $K_{sym}$ connected to the constant term, first (slope parameter) and second order (symmetry incompressibility) density derivative, respectively, all evaluated at $n_{sat}$. The prior
distribution for these aforementioned ones along with the others varied in this study is consistent with the present empirical knowledge for a
large set of nuclear data~\cite{Margueron:2017eqc}. The use of the
same functional to describe the core and the inhomogeneous
crust~\cite{Carreau:2019zdy} guarantees a consistent estimation of the crust--core transition inside the neutron star and, thus, consistent
predictions for its radius. This approach enables incorporating priors
from nuclear physics on the EoS and including the uncertainties at
high densities. The only limiting assumption is that matter is
composed of charged leptons, nucleons and nuclei only, and in
particular that no first--order phase transition occurs.  There might
also be some dependence on the different approximations applied to
treat inhomogeneous matter, \textit{e.g.} assuming vanishing temperature,
which could slightly modify the crust properties (see
\textit{e.g.}~\cite{Barba-Gonzalez:2022pkn}). However, this should not affect
the comparison of different configurations and only has a very small
impact on the extracted neutron star and nuclear matter
properties. The assumed nuclear prior complies with the chiral effective field theory
energy per particle band for symmetric and pure neutron matter as
given by \cite{Drischler:2015eba} for baryon number densities $ \SI{0.02}{\per\cubic\femto\meter}
\le n_B \le \SI{0.18}{\per\cubic\femto\meter}$, see \cite{Thi:2021jhz, Mondal:2022cva} for details
of the current implementation.

\subsection{Binary neutron star populations}

For ease of comparison, we use the same number of sources and redshift distribution as in \cite{Branchesi:2023mws} (to which we refer for further details), obtained with the population synthesis code \textsc{MOBSE}~\cite{Mapelli:2017hqk,Giacobbo:2017qhh}.\footnote{The catalog that we use has been provided by Michela Mapelli, and is publicly available at \url{https://apps.et-gw.eu/tds/?content=3&r=18321}.} In particular, the number of simulated sources in our \SI{1}{\year} catalogs is about \num{7.2e5}.
For the source--frame masses of the objects, $m_1^{\rm src}$ and $m_2^{\rm src}$, we choose uniform distributions between \SI{1.1}{\Msun} and the maximum allowed mass for each EoS. This allows us to study the impact of the EoS also on the number of detectable sources. From the two component masses we then obtain the quadrupolar adimensional tidal deformability parameters $\Lambda_{1}$ and $\Lambda_{2}$ through the $\Lambda(m)$ relation predicted by each EoS, reported in \autoref{fig:EoS_lamofm_plot}.

The aligned spin components, $\chi_{1,z}$ and $\chi_{2,z}$, are sampled from uniform distributions between $[-0.05,\,0.05]$,\footnote{The remaining spin components are set to $0$ and not included in the analysis. We do not expect this choice to strongly affect the results, given the small expected spins for neutron stars in binary systems.} and the remaining angular parameters (sky--position angles $\alpha$ and $\delta$, inclination $\iota$, polarization angle $\psi$, time of coalesce $t_c$ and phase at coalescence $\Phi_c$) are sampled uniformly in their physical range.

\subsection{Waveform models}
To study the impact of waveform modelling on our analysis, we use different waveforms with different recipes on how to treat the tidal contributions to the merger. In particular, we adopt 
\begin{description}[align=left]

    \item[\textsc{IMRPhenomD\_NRTidalv2}] \citep{Husa:2015iqa, Khan:2015jqa, Dietrich:2019kaq} frequency--domain phenomenological model built to describe the quadrupolar tidal contributions to the fundamental $(l=2,\,m=2)$ mode of a GW signal produced by a spin--aligned BNS coalescence;
    
    \item[\textsc{IMRPhenomD\_NRTidalv2\_Lorentzian}] \cite{Puecher:2022oiz} extension of the previous model, which further models the main emission peak of the post--merger phase of the signal; 

    \item[\textsc{SEOBNRv4T\_surrogate}] \cite{Bohe:2016gbl,Lackey:2018zvw} frequency--domain surrogate version of the aligned--spin BNS waveform effective one body (EOB) \cite{Buonanno:1998gg,Damour:2009wj,Hinderer:2016eia,Steinhoff:2016rfi} model \textsc{SEOBNRv4T} which includes both the quadrupolar and octupolar tidal contributions to the fundamental mode of the GW signal emitted by BNS systems.
    
\end{description}
For the EOB model, the tidal coefficients for the octupole are computed from quasi--universal relations in terms of the quadrupole coefficients using the relations provided in Ref.~\cite{Yagi:2016bkt}, thus keeping the number of parameters fixed.

\section{Fisher matrix framework}\label{sec:FIM_framework}
A fundamental building block of our analysis is the capability to forecast for each event in the catalogs its detectability and the attainable statistical uncertainties on its parameters. Given the size of the catalogs and the number of simulations needed, a full injection campaign and Bayesian parameter estimation is computationally unfeasible. We thus explore an approximated approach, in which the detectability of an event is determined by a threshold on its signal--to--noise ratio (SNR)
\begin{equation}\label{eq:SNRdefFull}
    {\rm SNR}^2 = 4 \Re{\int_{0}^{\infty} \dd{f} \dfrac{\tilde{h}^*(f)\tilde{h}(f)}{S_n(f)}}\,,
\end{equation}
where $\tilde{h}(f)$ denotes the Fourier--domain signal and $S_n(f)$ the detector noise power spectral density (PSD). The statistical errors on the source parameters for the detected events are then computed resorting to the Fisher matrix approximation, which is often used in the literature, and will be briefly reviewed in the following (see \cite{Cutler:1994ys,Vallisneri:2007ev,Rodriguez:2013mla} for comprehensive treatments). We stress that the performance of a real detector will obviously be different from the one simulated within this simplified framework, yet it can provide a sensible approximation that can be used to produce meaningful forecasts and, \textit{e.g.}, compare different detector configurations or choices for the population models. 
 
\subsection{Fisher formalism}
Assuming that the time--domain signal $s(t)$ in a GW detector can be expressed as the superposition of a signal $h(t;{\bm \theta}_0)$ (with ${\bm \theta}_0$ denoting the true parameters) and stationary, Gaussian noise $n(t)$ with zero mean, \textit{i.e.} $s(t)=h(t;{\bm \theta}_0)+n(t)$, the likelihood for a data realisation $s(t)$ conditioned on the parameters ${\bm \theta}$ of a waveform template is given by
\begin{equation}\label{eq:gwlikFull}
\mathcal{L}(s \,|\, {\bm \theta}) \propto \exp{-\dfrac{1}{2}\left( s -h({\bm \theta})  \, | \, s -h({\bm \theta}) \right)} \, ,
\end{equation}
with the inner product $(a\,|\,b)$ being defined as
\begin{equation}\label{eq:inner_prod_def}
    ( a \, | \, b ) \equiv  4 \Re{ \int_0^{\infty} \dd{f} \, \frac{\tilde a^*(f) \, \tilde b(f) }{S_{n}(f)}} \, ,
\end{equation}
where the tilde denotes a temporal Fourier transform. From this definition it follows that \autoref{eq:SNRdefFull} can be expressed as ${\rm SNR} = \left[(h\,|\,h)\right]^{\nicefrac{1}{2}}\,.$ 

Expanding the template signal around the true values of the parameters ${\bm \theta}={\bm \theta}_0$ and retaining only first derivatives of the signal (this goes under the name of \emph{linearized signal approximation}, LSA, which is equivalent to the high--SNR limit \cite{Vallisneri:2007ev}), the likelihood in \autoref{eq:gwlikFull} reduces to a multivariate Gaussian. In what follows, we will also focus on the limit of zero noise, in which case the explicit expression for the LSA likelihood is 
\begin{equation}\label{eq:LSAgwlik}
\mathcal{L}(s \,|\, {\bm \theta}) \propto \exp \left\{-\dfrac{1}{2} \delta\theta^{i}\delta\theta^{j} \Gamma_{ij} \right\} \, ,
\end{equation}
where $\bm{\delta\theta} \equiv {\bm \theta} - {\bm \theta}_0$ and we have introduced the Fisher information matrix (FIM)
\begin{equation}\label{eq:FIMdef}
\Gamma_{ij} \equiv \left.\left(\pdv{h}{\theta^i}\,\right|\left.\pdv{h}{\theta^j}\right)\right|_{{\bm \theta} = {\bm \theta}_0}\,.
\end{equation}
The inverse of the FIM thus gives the covariance matrix of the likelihood in \autoref{eq:LSAgwlik}, ${\rm Cov}_{ij}=\Gamma_{ij}^{-1}$, from which we can get the statistical errors on the template parameters as $\sigma_i = \sqrt{{\rm Cov}_{ii}}$.

In our analysis, the parameters used to characterise the GW signal are 
\begin{equation}
    {\bm \theta} = \{{\cal M}_c,\, \eta,\, d_L,\, \alpha,\, \delta,\,\iota,\,\psi,\, t_c,\, \Phi_c,\, \chi_{1,z},\, \chi_{2,z},\, \tilde{\Lambda},\,\delta\tilde{\Lambda}\}\,,
\end{equation}
where ${\cal M}_c$ denotes the detector--frame chirp mass, $\eta$ the symmetric mass ratio, and the combinations of the quadrupolar adimensional tidal deformabilities  $\tilde{\Lambda}$ and $\delta\tilde{\Lambda}$ are given by~\cite{Wade:2014vqa} 
\begin{equation}\label{eq:tildeLam_def}
    \begin{aligned}
        \tilde{\Lambda} &= \dfrac{8}{13} \left[(1+7\eta-31\eta^2)(\Lambda_1 + \Lambda_2) \right.\\
        & \left. + \sqrt{1-4\eta}(1+9\eta-11\eta^2)(\Lambda_1 - \Lambda_2)\right]\,;\\
        \delta\tilde{\Lambda} &= \dfrac{1}{2} \left[\sqrt{1-4\eta} \left(1-\dfrac{13272}{1319}\eta + \dfrac{8944}{1319}\eta^2\right)(\Lambda_1 + \Lambda_2) \right.\\
        & \left.+ \left(1 - \dfrac{15910}{1319}\eta + \dfrac{32850}{1319}\eta^2 + \dfrac{3380}{1319}\eta^3\right)(\Lambda_1 - \Lambda_2)\right]\,.
    \end{aligned}
\end{equation}
To estimate SNRs and FIMs we use the publicly available code named \textsc{GWFAST} which was developed recently \cite{Iacovelli:2022mbg,Iacovelli:2022bbs}.\footnote{\textsc{GWFAST} is available at \url{https://github.com/CosmoStatGW/gwfast}.}\textsuperscript{,}\footnote{
 For other recent Fisher parameter estimation codes see~\cite{Borhanian:2020ypi,Borhanian:2022czq, Dupletsa:2022wke,Chan:2018csa,Li:2021mbo,Pieroni:2022bbh}. Results from these codes were cross--checked within the activities of the ET Observational Science Board~\cite{Branchesi:2023mws, Iacovelli:2022bbs}.
} 

\subsection{Checks with full parameter estimation runs}
\label{subsec:FIM_PE_comparison}

\begin{table*}[t]
    \centering
    \begin{tabular}{||c|c|c|c|c|c|c|c|c|c||}
    \toprule\midrule
    Source & $m_1\,[{\rm M}_{\odot}]$ & $m_2\,[{\rm M}_{\odot}]$ & $d_L\,[{\rm Mpc}]$ & $\chi_{1,z}$ & $\chi_{2,z}$ & $\Lambda_{1}$ & $\Lambda_{2}$ & SNR (\textsc{Phenom}) & SNR (\textsc{SEOBNR})\\
    \midrule
    1 & 1.35 & 1.34 & 100 & 0.02 & 0.03 & 275 & 309 & 570.68 & 570.63\\
    2 & 1.42 & 1.18 & 100 & $-0.03$ & 0.04 & 276 & 898 & 134.99 & 134.98\\
    3 & 1.95 & 1.88 & 360 & $0.02$ & 0.04 & 18 & 27 & 90.88 & 90.87\\
    4 & 1.80 & 1.67 & 460 & $0.0$ & $-0.03$ & 41 & 83 & 79.10 & 79.08\\
    \midrule\bottomrule
    \end{tabular}
    \caption{Table reporting the parameters of the sources used for the comparison among Fisher and full PE results. The masses are given in source frame. In the last two columns we report the SNR for each source (in the 2L configuration) obtained with the \textsc{IMRPhenomD\_NRTidalv2} and \textsc{SEOBNRv4T\_surrogate} approximants, respectively.}
    \label{tab:FishvsPE_evs}
\end{table*}

\begin{table}[b]
	\setlength\extrarowheight{4pt}
	\begin{tabular}{||c|c||}
		\toprule\midrule
		Parameter	& Range  \\ [0.5ex]
		\midrule
		$\mathcal{M}_c\,[\si{\Msun}]$ & [$\mathcal{M}_{c,s} \pm 0.05$ ] \\
		$q$ & [0.5, 1] \\
		$\chi_1$, $\chi_2$ & [0, 0.15] \\
        $d_L^{(1,2)}\,[\si{\mega\parsec}]$ & [1,500] \\
		$d_L^{(3,4)}\,[\si{\mega\parsec}]$ & [1,750] \\
		$\Lambda_1$, $\Lambda_2$ & [0, 5000] \\
		\midrule\bottomrule
	\end{tabular}
	\caption{Priors employed in the PE analysis, where $\mathcal{M}_{c,s}$ represents the chirp mass injected value of the specific source analyzed. For the luminosity distance $d_L$, the prior is taken uniform in comoving volume; for all the other parameters, the prior is uniform in the indicated range. The $d_L$ prior for Source 3 and Source 4 is wider because of the higher injected values of $d_L$.}
	\label{tab:priors}
\end{table}

Before proceeding with a full population analysis and EoS
reconstruction, we want to assess the reliability of the Fisher matrix
approach in the estimation of statistical uncertainties on the masses
and tidal deformability parameters of BNS systems. To this purpose, we perform a
comparison on some selected events of the FIM results with full
Bayesian parameter estimation (PE) runs. We simulate
signals with the parameters given in \autoref{tab:FishvsPE_evs} in
zero noise, in order to avoid biases caused by noise fluctuations, and to be able to consistently compare with the results of the Fisher analysis.

We employ the injection framework available
in the \textsc{Bilby} library
\cite{Ashton:2018jfp,Romero-Shaw:2020owr}.  We simulate signals with
the \textsc{IMRPhenomD\_NRTidalv2} and \textsc{SEOBNRv4T\_surrogate}
waveforms, and we recover them with the same model used for
injections.  For the parameter estimation analysis, the likelihood is
sampled with the nested sampling \cite{Skilling:2006gxv,Veitch:2009hd}
algorithm \textsc{dynesty}
\cite{Speagle:2019ivv,sergey_koposov_2022_6456387} included in
\textsc{Bilby}. We analyze the data starting
from 10~Hz, since going to lower frequencies would significantly
increase the computational cost of the analysis.\footnote{When performing these checks, for consistency, we also compute the FIMs with a frequency grid starting at \SI{10}{\hertz}, while in the following analyses we use a low--frequency cutoff of \SI{2}{\hertz}.} The computational
cost of PE analysis increases with both the signal duration and the
frequency range we want to study. The BNS signals considered for this
comparison study have a duration of roughly 20 minutes, and choosing
lower values for the starting frequencies adds a significant amount of points to
the frequency grid over which we need to evaluate the waveforms for
the likelihood computation. In order to make this analysis
computationally feasible, we employ the relative binning technique
\cite{Zackay:2018qdy, Leslie:2021ssu}, as implemented in
\cite{janquart:2022}. Moreover, to further reduce the computational cost, we
here focus on the 2L configuration only.

\begin{figure*}[htp!]
    \centering
    \includegraphics[width=.7\textwidth]{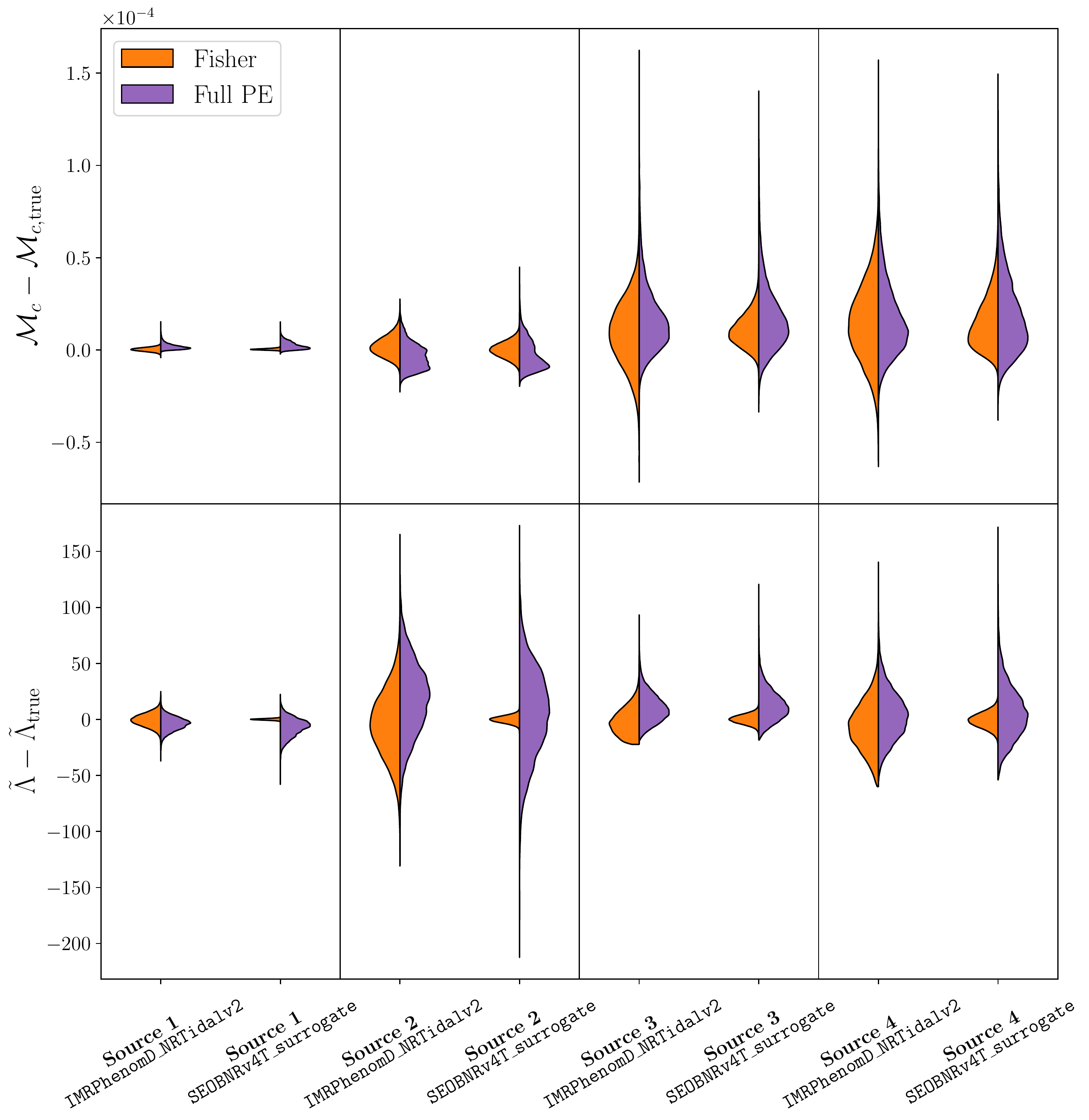}
    \caption{Violin plots comparing the measurement errors results for the chirp mass (top row) and $\tilde{\Lambda}$ parameter (bottom row) obtained for the different considered sources in \autoref{tab:FishvsPE_evs} and the two waveforms \textsc{IMRPhenomD\_NRTidalv2} and \textsc{SEOBNRv4T\_surrogate} adopting the Fisher formalism (orange) and performing a full Bayesian parameter estimation (violet). For ease of readability and comparison among different sources, we report the distribution of ${\cal M}_c$ and $\tilde{\Lambda}$ minus the injected value.}
    \label{fig:AllSources_FIMvsPE_LambdaTilde}
\end{figure*}

The parameters of the selected sources are listed in \autoref{tab:FishvsPE_evs},\footnote{The difference in SNR among the first and second source, which are at the same luminosity distance, is due to a difference in the other angular parameters used for the injection.} and the priors employed are listed in \autoref{tab:priors}. In \autoref{fig:AllSources_FIMvsPE_LambdaTilde}, we report violin plots comparing the FIM estimations and the full PE runs for ${\cal M}_c$ and the $\tilde{\Lambda}$ parameter (we here impose a prior $m_1\geq m_2$ in the FIM results, which generates the small observed non--Gaussianities). We overall find a good level of agreement among FIM and full PE on the error estimate for $\tilde{\Lambda}$ (we here focus on the width of the distributions) when using the \textsc{IMRPhenomD\_NRTidalv2} waveform model. However, this does not hold for \textsc{SEOBNRv4T\_surrogate}, in which case the FIM results underestimate the error attainable on $\tilde{\Lambda}$, while the estimation is consistent among the full PE runs with different waveform models. The disagreement is much less pronounced on the chirp mass estimation. This behavior can be understood from the variation produced in the output of a waveform model as a consequence of a change in a parameter. We can quantify this variation from the \emph{mismatch}, $\mathscr{M}$, among the waveform predicted by an approximant for a given set of parameters and
the prediction of the same model when varying one of the
parameters. In particular, given the overlap between two signals
\begin{equation}
    \mathscr{O}(h_1, h_2) = \dfrac{(h_1|h_2)}{\sqrt{(h_1|h_1)(h_2|h_2)}}\,,
\end{equation}
where $(\vdot|\vdot)$ denotes the inner product defined in \autoref{eq:inner_prod_def}, we quantify the mismatch in two different ways: a `direct' estimation, given by
\begin{equation}\label{eq:match_direct}
    \mathscr{M}_{d}(h_1, h_2) = 1 - \mathscr{O}(h_1, h_2)\,,
\end{equation}
and the `standard' definition (see \textit{e.g.} \cite{Owen:1995tm, Ohme:2011zm})
\begin{equation}\label{eq:match_usual}
    \mathscr{M}_{s}(h_1, h_2) = 1 - \max_{t_{c}, \Phi_{c}}\mathscr{O}(h_1, h_2)\,.
\end{equation}
The advantage of the first definition is that the computation of
$\mathscr{M}_{d}$ for a chosen template varying a single parameter is
closely related to the Fisher matrix, which is built upon the signal
partial derivatives with respect to each parameter (keeping all the
others fixed). $\mathscr{M}_{d}$ is thus useful to have some insight
on the FIM behaviour. 

\begin{figure*}[tp]
    \centering
    \includegraphics[width=.9\textwidth]{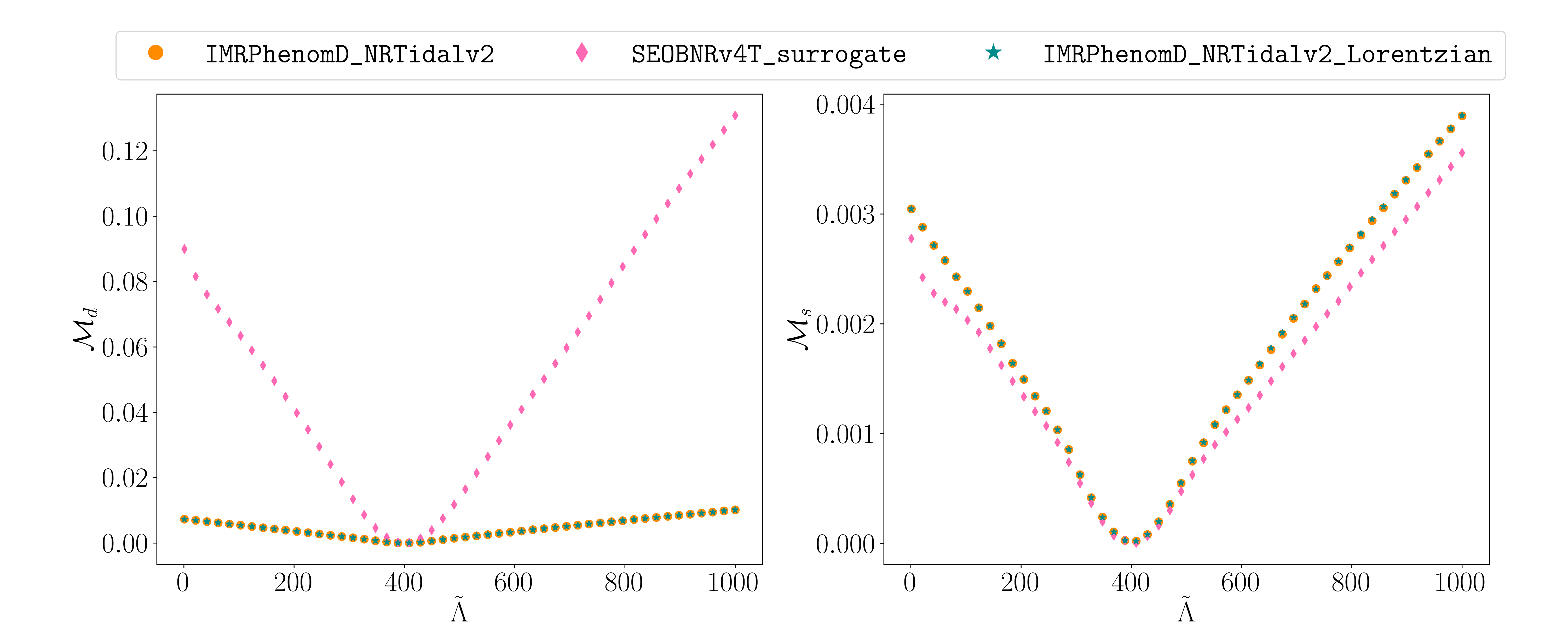}
    \caption{Mismatch of the various templates considered for the first source in \autoref{tab:FishvsPE_evs} varying the $\tilde{\Lambda}$ parameter, using: the definition in \autoref{eq:match_direct} (left panel) and the one in \autoref{eq:match_usual} (right panel).
    }
    \label{fig:mismatch_source1}
\end{figure*}

In \autoref{fig:mismatch_source1}, we report the
mismatch for the source 1 estimated both using the definition in
\autoref{eq:match_direct} (left panel) and in \autoref{eq:match_usual}
(right panel). Similar behaviours are observed also for the other
events considered in this section.

As it is apparent from the plots in \autoref{fig:mismatch_source1}, the variation in the \textsc{SEOBNRv4T\_surrogate} model is more pronounced than for \textsc{IMRPhenomD\_NRTidalv2} when using the mismatch definition in \autoref{eq:match_direct}. Thus, when taking derivatives keeping all the parameters fixed but the tidal deformability $\tilde{\Lambda}$, the former approximant will result in a bigger FIM element for this parameter, and consequently a smaller error when inverting the FIM to obtain the covariance matrix. On the other hand, if the mismatch is computed by maximizing the overlap over the coalescence time and phase, as defined in \autoref{eq:match_usual}, we find comparable trends for \textsc{SEOBNRv4T\_surrogate} and \textsc{IMRPhenomD\_NRTidalv2}. This reflects the results we obtain for full PE runs, where we sample also on time and phase, and then marginalize over them, together with the other parameters, in order to find the $\tilde{\Lambda}$ posterior.

Regarding the Fisher matrix approach, we further verified that the uncertainty estimates on $\tilde{\Lambda}$ obtained using \textsc{SEOBNRv4T\_surrogate} are consistently smaller than the ones obtained from \textsc{IMRPhenomD\_NRTidalv2}, with differences as big as one order of magnitude on average, irrespectively of the chosen EoS model or detector configuration. This also holds at the level of population study. Also in this case, the effect is considerably less pronounced for the uncertainties on the masses. 
Given the findings reported in this section, in the following analyses we will not employ the \textsc{SEOBNRv4T\_surrogate} model.\footnote{We also performed all the same checks employing the \textsc{TEOBResumSPA} frequency--domain EOB approximant \cite{Nagar:2018zoe,Nagar:2018plt,Nagar:2019wds,Riemenschneider:2021ppj,Gamba:2020ljo}, which can include the quadrupolar, octupolar and hexadecapolar tidal contributions to the signal, as well as subdominant harmonics, finding, in the FIM case, error estimates even tighter than the ones obtained employing \textsc{SEOBNRv4T\_surrogate}, both on single sources and at the population level. We did not perform full PE runs for this model due to technical issues with the relative binning method employed here, but, given the results obtained for \textsc{SEOBNRv4T\_surrogate}, we consider the FIM results for the reconstruction of the tidal parameters too optimistic also in this case. 
For these reasons, we decided not to employ this model either in the following.} We stress again that the observed behaviour is present only in the Fisher forecasts, and the full PE results obtained in our simulations with the two models are consistent. 

\section{Results}\label{sec:results}
We first summarise our main findings here regarding the
observational prospects of BNS systems in  ET depending on the
underlying EoS, along with the statistical uncertainties attainable on
their tidal deformability parameters. We then focus on the
reconstruction of the underlying NS properties and the NS EoS based on these
observations, exploring, in addition, the attainable accuracy on a set
of independent empirical parameters characterising the density
dependence of the energy in symmetric matter and of the symmetry
energy.

\subsection{Population analysis results}\label{subssec:pop_results}
\begin{figure*}[htb!]
\centering
\begin{tabular}{l@{\hskip 1.cm}l}
     \includegraphics[width=7.5cm]{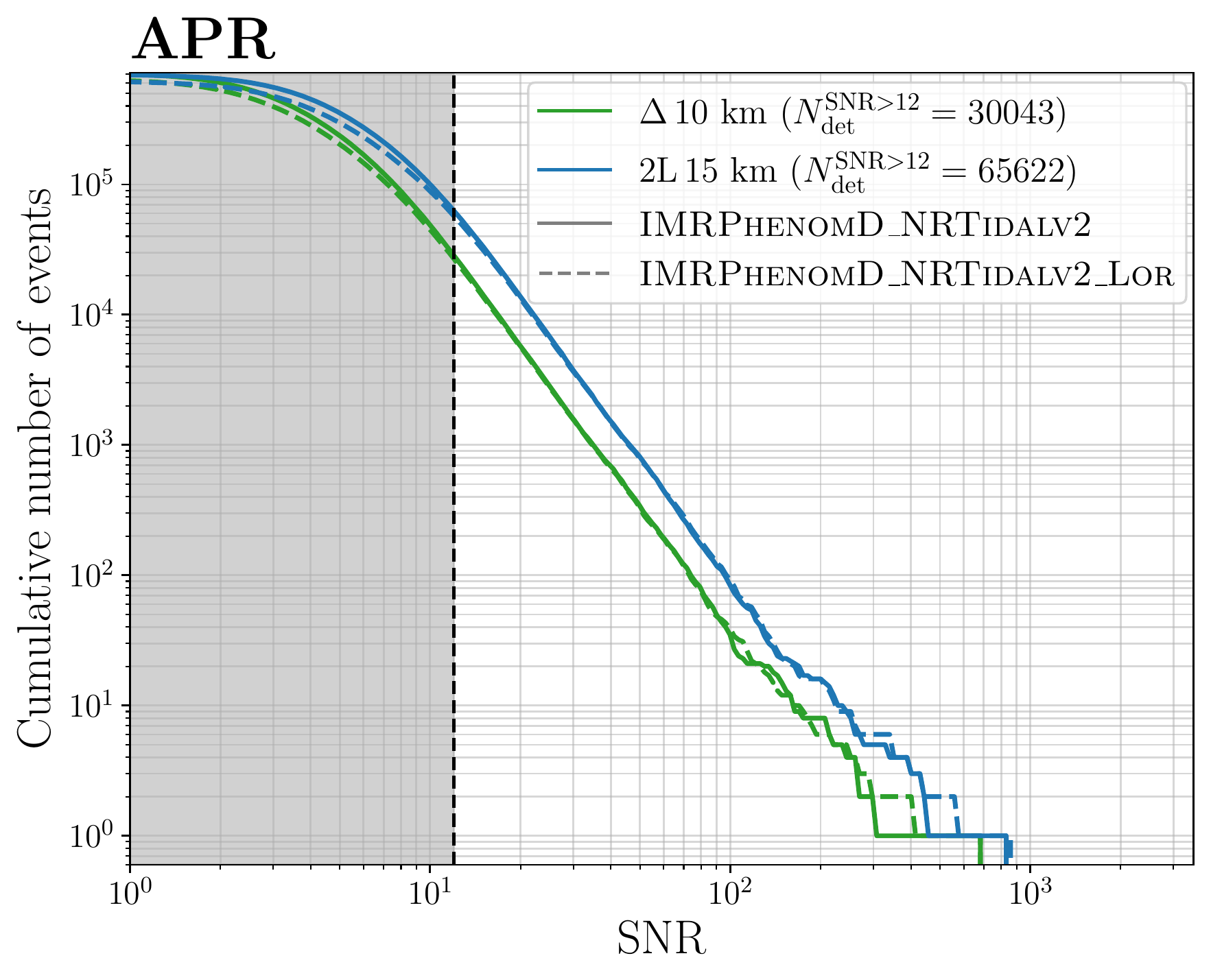} & 
     \includegraphics[width=7.5cm]{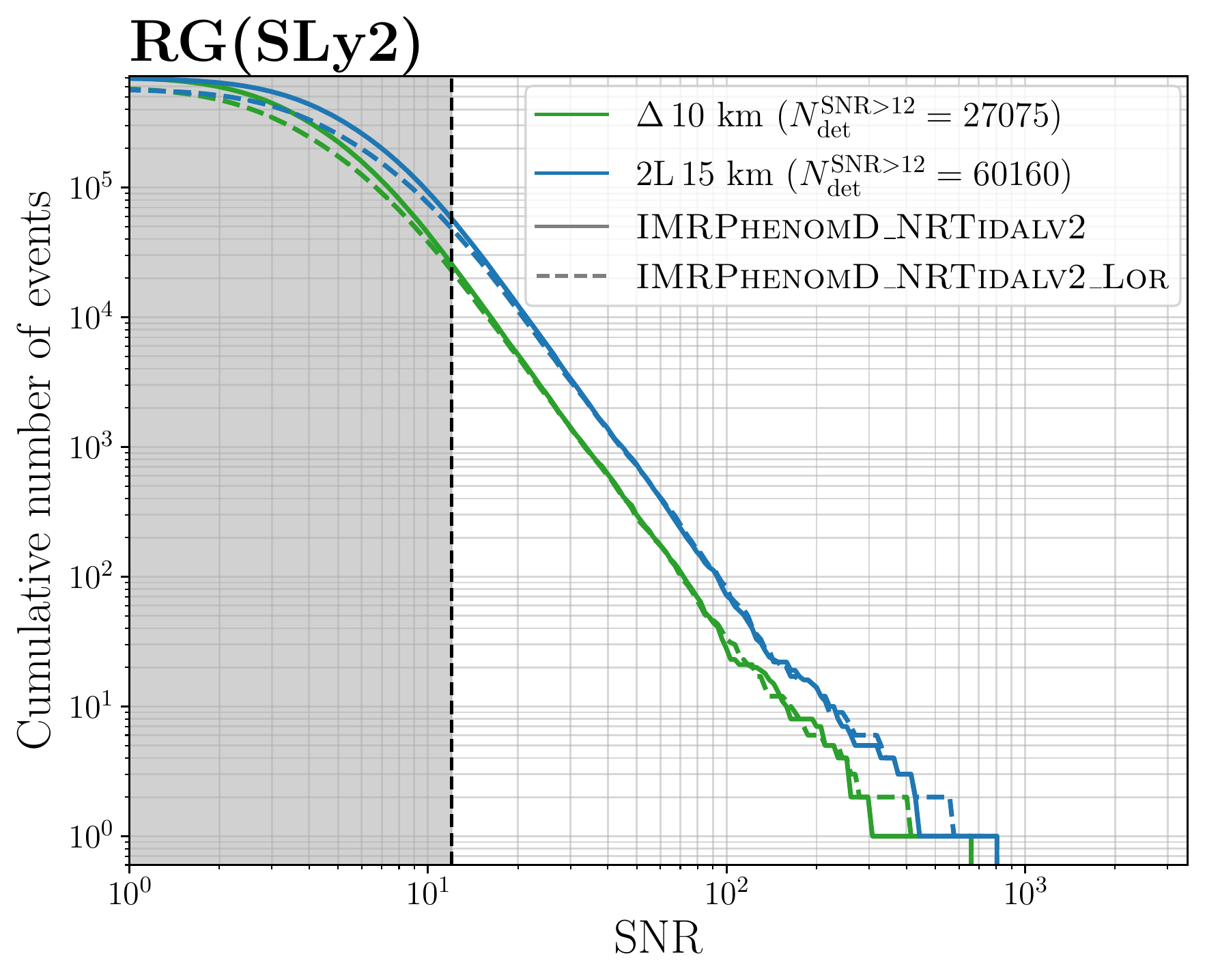} \\
     \includegraphics[width=7.5cm]{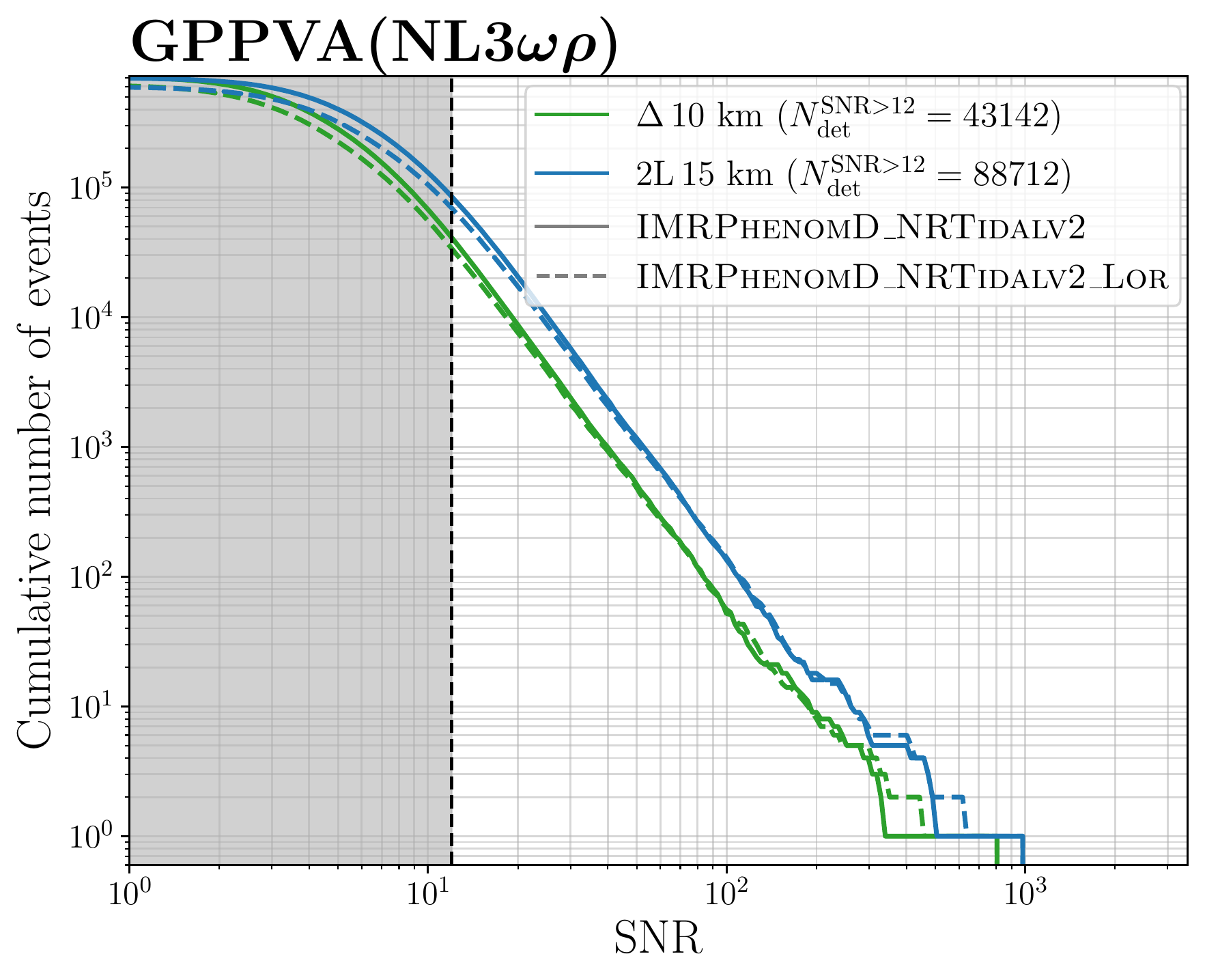} & 
     \includegraphics[width=7.5cm]{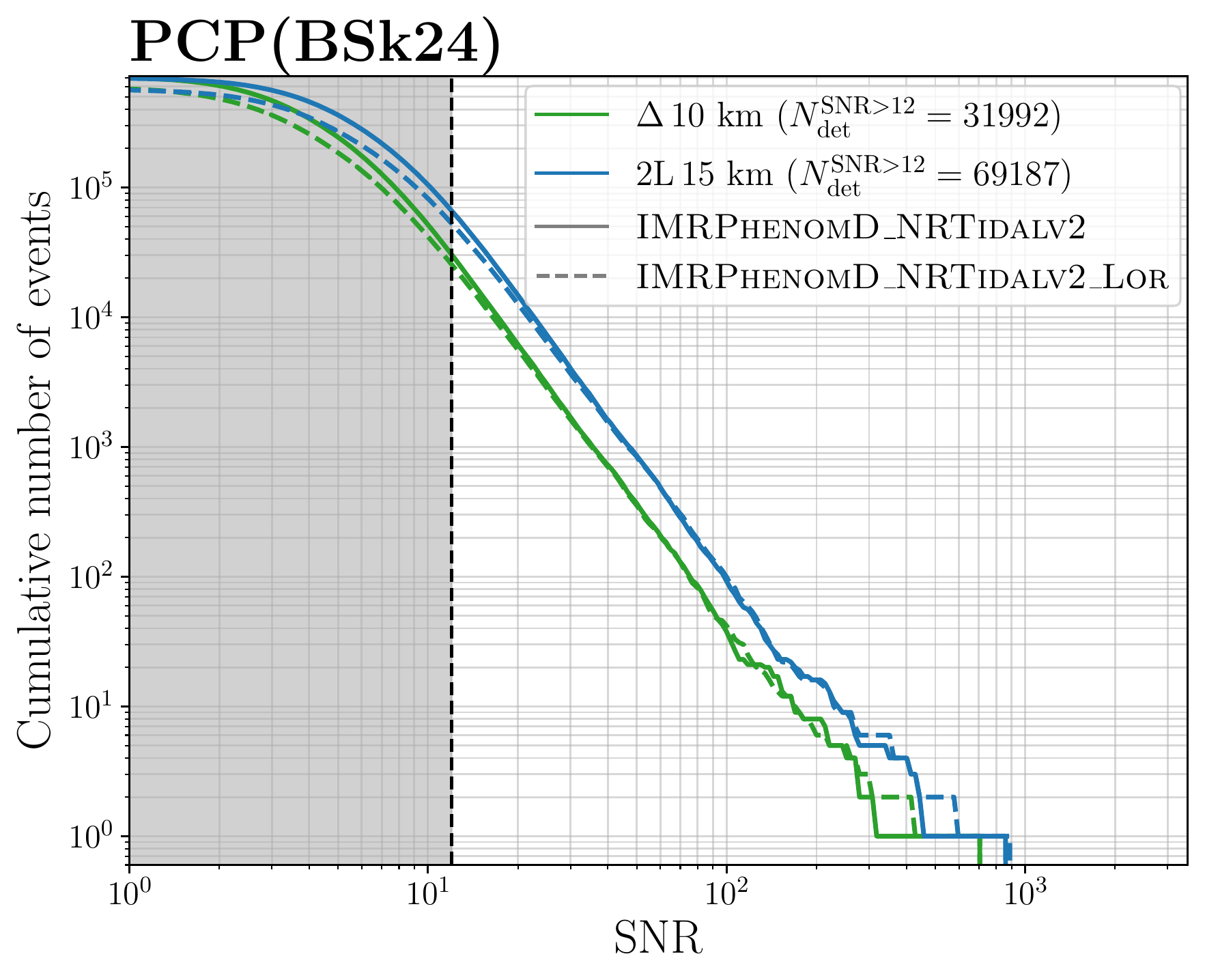}
\end{tabular}
    \caption{(Inverse) cumulative distributions of the SNRs for the considered population as observed by ET in the triangular geometry (green lines) and 2L-$45^\circ$ geometry (blue lines) for the 2 considered waveform models \textsc{IMRPhenomD\_NRTidalv2} (solid) and \textsc{IMRPhenomD\_NRTidalv2\_Lorentzian} (dashed). Each panel shows the forecasts obtained adopting different EoS models, reported in the title. The shaded area denotes the region below SNR=12, which we use as detection threshold. In the legend we further report the number of observed events with ${\rm SNR}\geq12$ with the \textsc{IMRPhenomD\_NRTidalv2} approximant.}
    \label{fig:AllEos_SNR_pop_cumul}
\end{figure*}

In \autoref{fig:AllEos_SNR_pop_cumul}, we report cumulative distributions of the SNRs for the considered populations and network configurations. In each run we assume an uncorrelated 85\% duty cycle for each interferometer. Overall, we find a strong dependence of the number of detections on the underlying EoS. This can be traced to the difference in the maximum masses predicted by the different equations of state: the ones allowing higher masses result in higher number of detections, since more massive systems produce a louder signal. Setting an SNR threshold of 12 for a signal to be detectable, we find that the number of observable sources (reported in the legend of \autoref{fig:AllEos_SNR_pop_cumul}) can vary by a factor $\ssim1.5$ comparing the results obtained with RG(SLy2), which predicts the smallest $m_{\rm max}$ (cf.~legend of \autoref{fig:EoS_lamofm_plot}) among the chosen models, and the ones obtained with GPPVA(NL3$\omega\rho$), which predicts the highest one.\footnote{\label{footnote:massdistr_popres} This also follows from our choice of a flat mass distribution up to the maximum allowed mass: we verified that, sampling independently the two components' masses from a Gaussian distribution with mean $1.33~{\rm M}_{\odot}$ and standard deviation $0.09~{\rm M}_{\odot}$ \cite{Farrow:2019xnc} [always truncated at the maximum mass allowed by each EoS], the number of detections is overall lower as compared to the flat distribution case, due to the lack of more massive objects, but rather similar among different EoS models ($\ssim1.7\times10^4\,{\rm events/yr}$ for the triangular design and $\ssim4.2\times10^4\,{\rm events/yr}$ for the 2L configuration). This can be understood from the fact that the mass cut imposed by each EoS is many sigmas away from the mean of the chosen distribution.} We notice the overall higher number of detections obtained with the two L--shaped interferometers of 15~km as compared to the triangular design with 10~km long arms. Nevertheless, we find the number of possible detections to be never lower than $\ssim2.7\times10^4\, {\rm events/yr}$ with our assumptions on the merger rate and  mass distribution. 

\begin{figure*}[tbp]
\centering
\begin{tabular}{l@{\hskip 1.cm}l}
     \includegraphics[width=7.5cm]{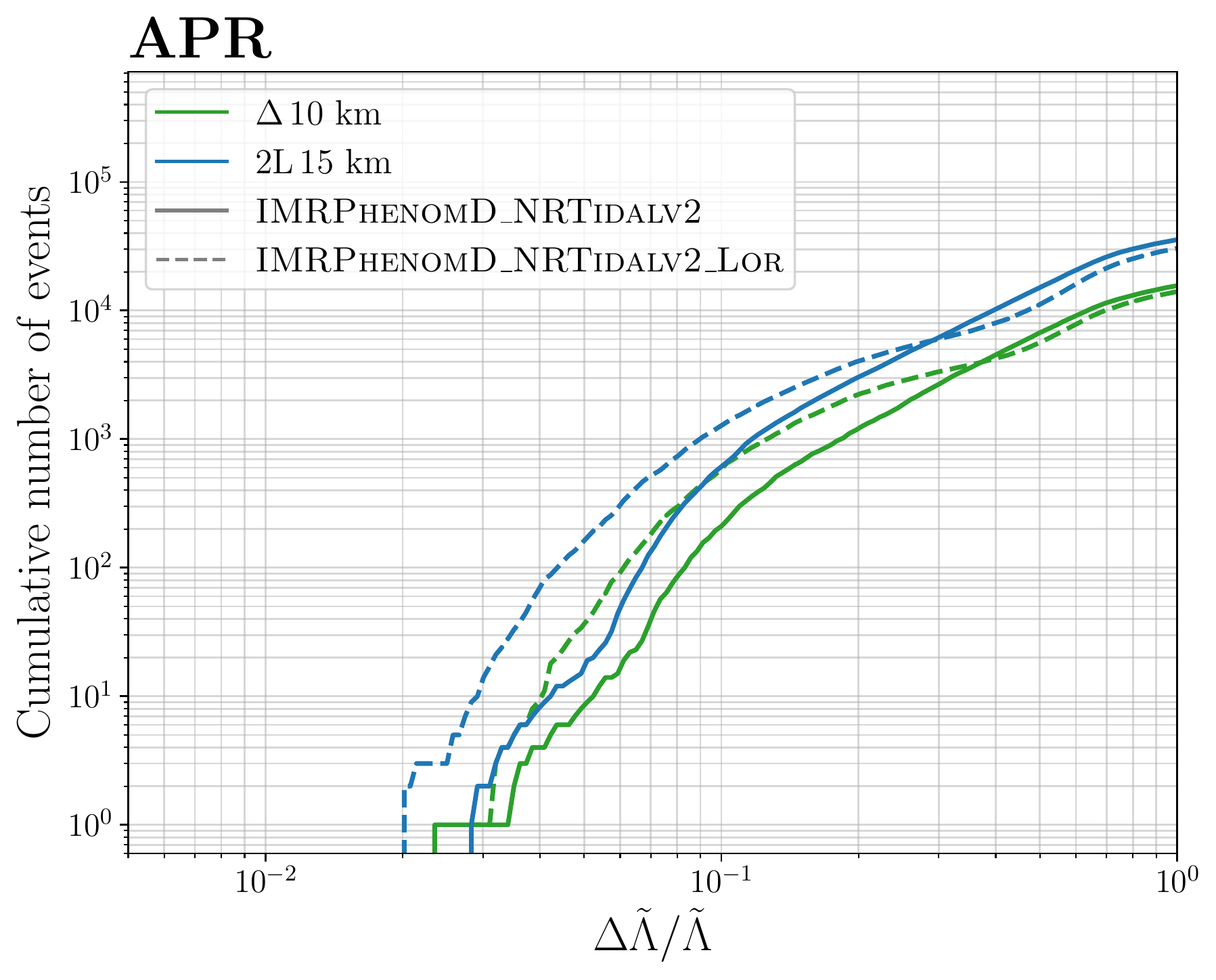} & 
     \includegraphics[width=7.5cm]{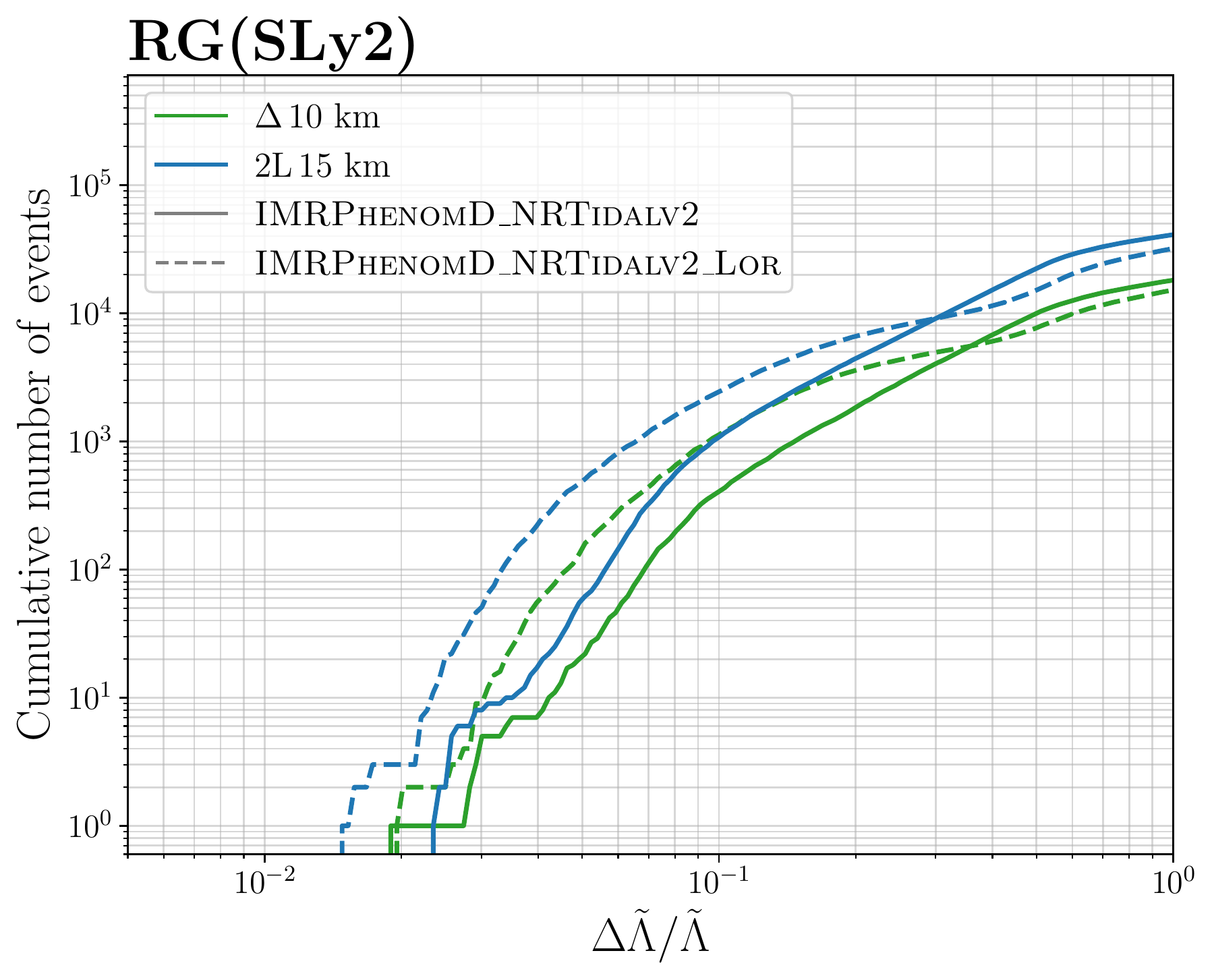} \\
     \includegraphics[width=7.5cm]{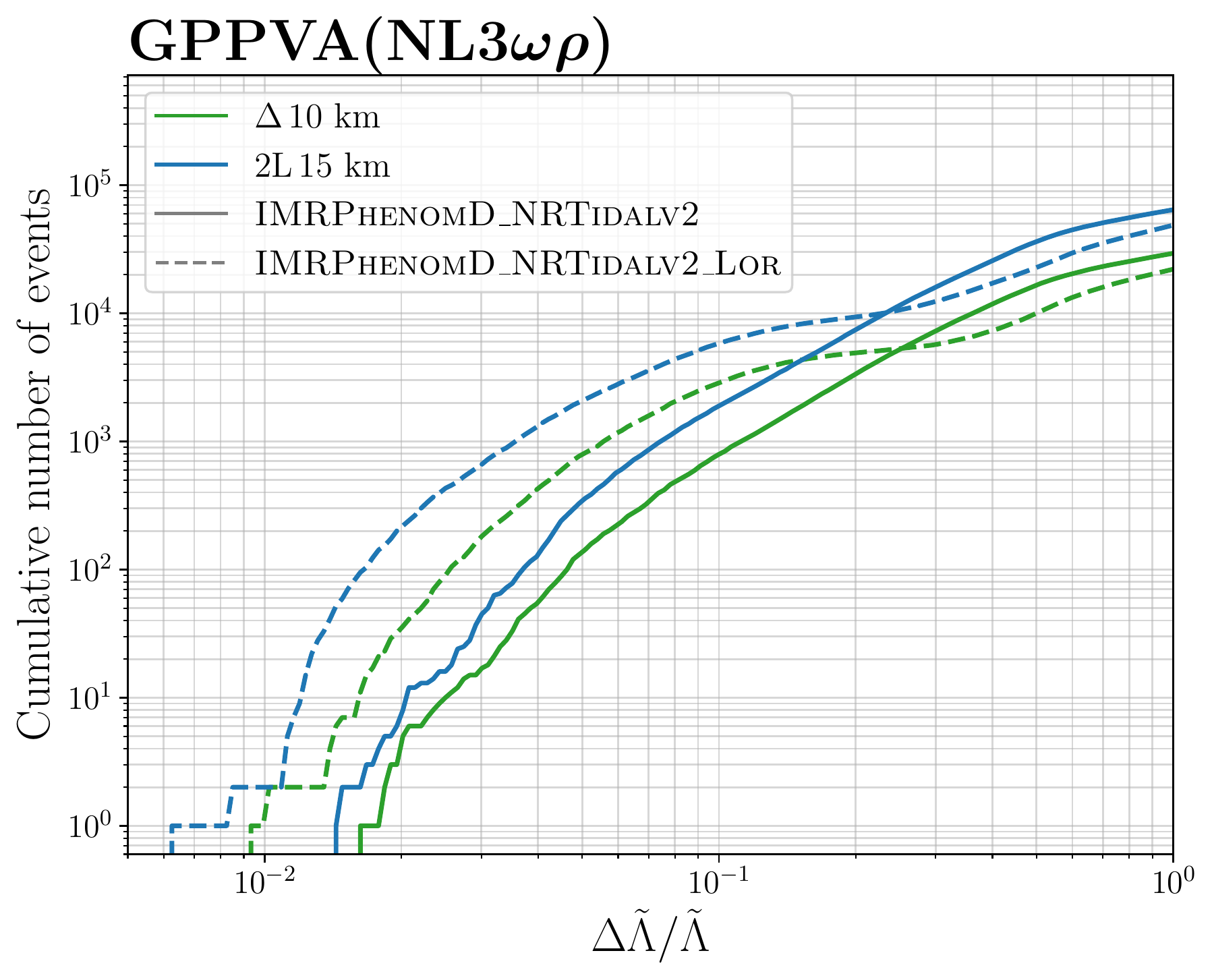} & 
     \includegraphics[width=7.5cm]{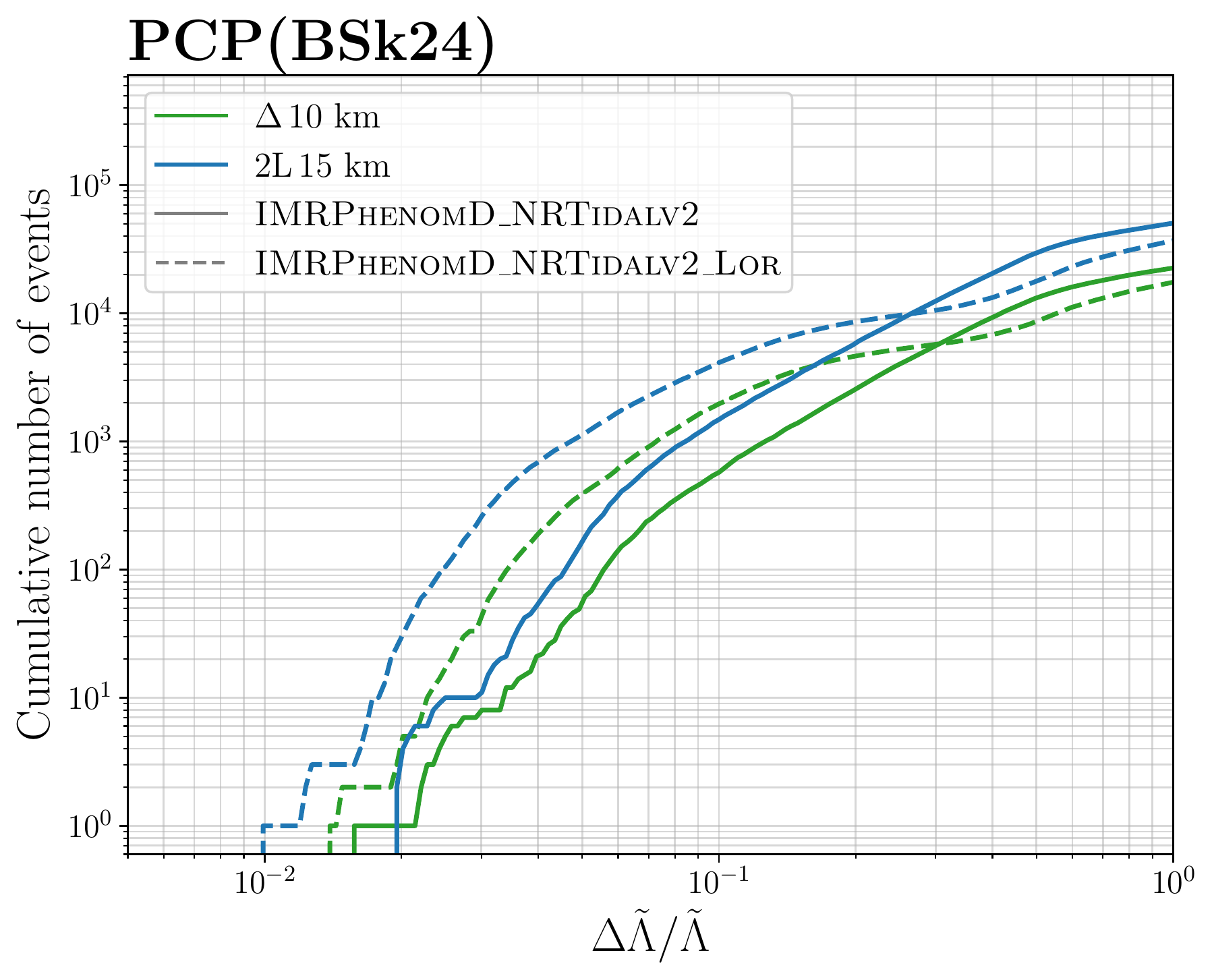}    
\end{tabular}
    \caption{Cumulative distributions of the relative 1-$\sigma$ relative statistical uncertainties attainable on the adimensional tidal deformability combination $\tilde{\Lambda}$ for the considered population as observed by ET in the triangular geometry (green lines) and 2L-$45^\circ$ geometry (blue lines) for the two considered waveform models \textsc{IMRPhenomD\_NRTidalv2} (solid) and \textsc{IMRPhenomD\_NRTidalv2\_Lorentzian} (dashed). Each panel shows the forecasts obtained adopting different EoS models, reported in the title.}
    \label{fig:AllEos_DelLam_pop_cumul}
\end{figure*}

In \autoref{fig:AllEos_DelLam_pop_cumul}, we report cumulative distributions of the relative 1-$\sigma$ statistical uncertainties attainable on $\tilde{\Lambda}$ for all the events that pass the chosen detection criteria ${\rm SNR}\geq12$.\footnote{We further discard a small fraction ($\lesssim1\%$) of events with unreliable inversion of the FIM, setting a threshold on the inversion error $\epsilon\equiv\Gamma\vdot{\rm Cov}\leq0.05$, as done in \cite{Iacovelli:2022bbs}. The same cuts in SNR and inversion error are employed in all the subsequent analyses.} From these plots we see again the overall better performance of the 2L geometry, but we also notice that the level of accuracy attainable on the very best events is comparable among the two configurations, always being at the few percent level for the \textsc{IMRPhenomD\_NRTidalv2} model. We further appreciate the gain in terms of accuracy that can be obtained on $\tilde{\Lambda}$ thanks to the modelling of the post--merger phase present in the \textsc{IMRPhenomD\_NRTidalv2\_Lorentzian}.

\subsection{Neutron star properties}\label{subssec:NS_properties_results}
\begin{figure*}[htbp]
    \centering
    \includegraphics[width=.94\textwidth]{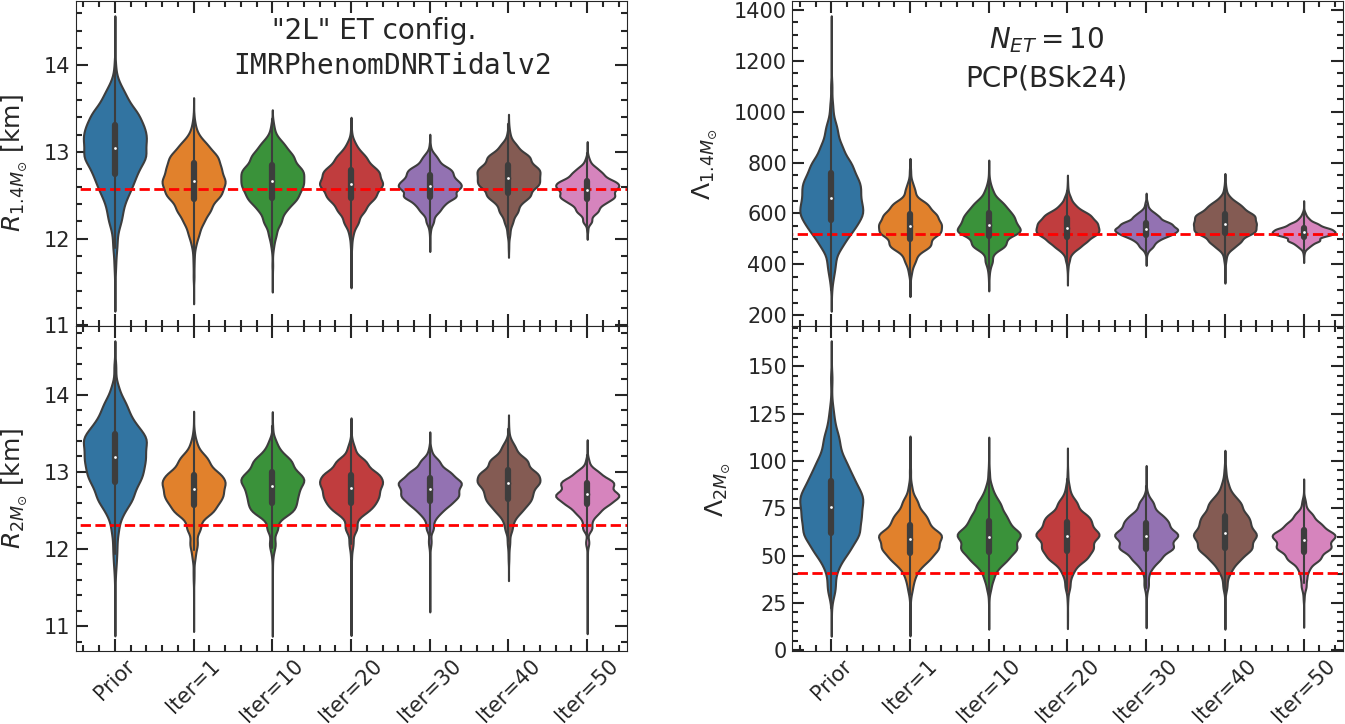}
    \caption{NS properties for $N_{\rm ET}=10$ over 50 iterations obtained for the 2L design with the PCP(BSk24) EoS.}
    \label{fig:NS_iter}
\end{figure*}
\begin{figure*}[htbp]
    \centering
    \includegraphics[width=.94\textwidth]{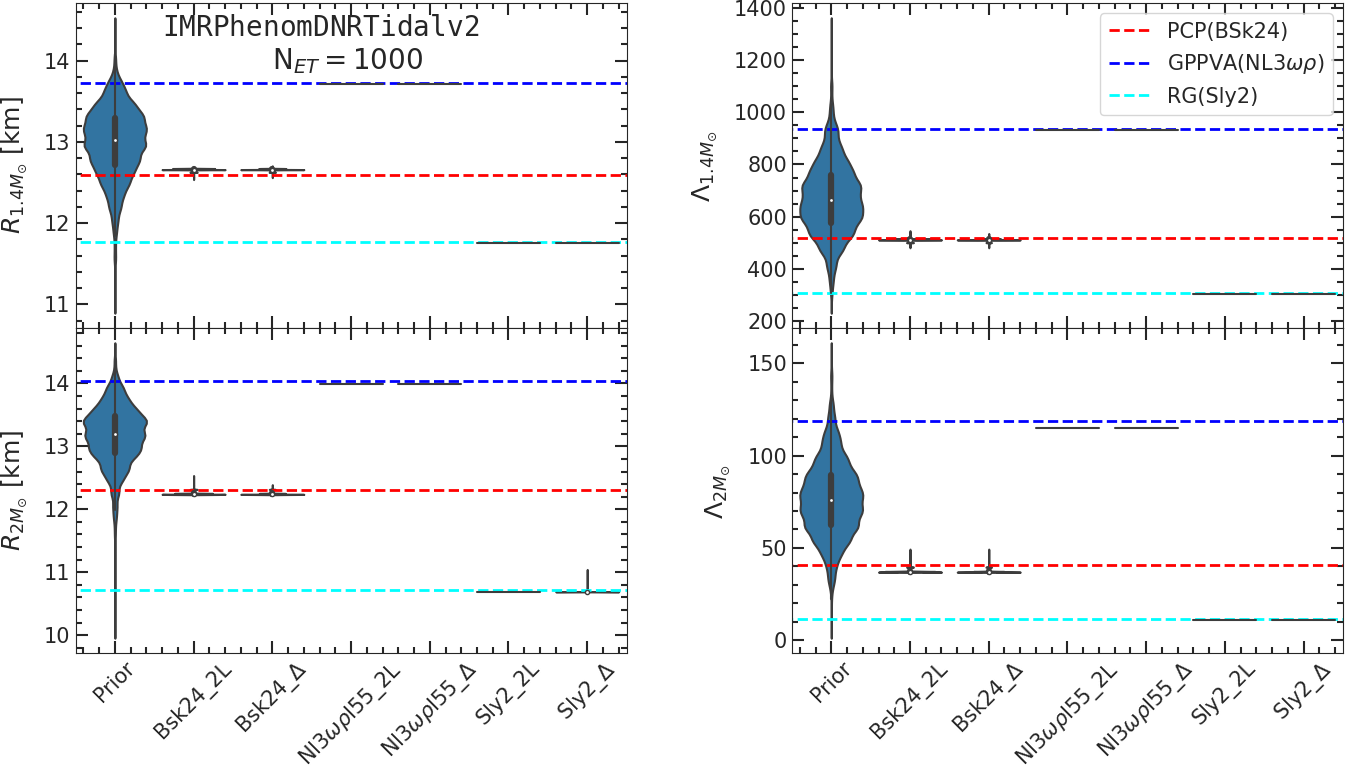}
    \caption{Selected NS properties for detector configurations 2L and $\Delta$ obtained with $N_{\rm ET}=1000$ and three different injected EoS models (see legend). The obtained values are listed in \autoref{tab:NS_radius_eos_detector} and \ref{tab:NS_lambda_eos_detector}, respectively.}
    \label{fig:NS_detector}
\end{figure*}
\begin{table*}[tbh]
	\setlength\extrarowheight{4pt}
	\begin{tabular}{||c|c|c|c|c|c|c||}
		\toprule\midrule
		
		EoS &  \multicolumn{3}{c|}{$R_{1.4{\rm M}_{\odot}}$ [km]} &  \multicolumn{3}{c|}{$R_{2{\rm M}_{\odot}}$ [km]}\\ [0.5ex]
		\midrule
  & Injected & 2L \SI{15}{\kilo\meter} & ‌$\Delta$ \SI{10}{\kilo\meter} & Injected & 2L \SI{15}{\kilo\meter} & $\Delta$ \SI{10}{\kilo\meter}\\
  \midrule
  PCP(BSk24) & 12.59 & $12.660^{+0.011}_{-0.001}$ & $12.662^{+0.010}_{-0.003}$ & 12.31 & $12.246^{+0.006}_{-0.008}$ & $12.249^{+0.003}_{-0.011}$ \\
		\midrule
  GPPVA(NL3$\omega\rho$) & 13.73 & $13.714^{+5e-9}_{-5e-9}$ & $13.714^{+5e-9}_{-5e-9}$ & 14.04 & $13.989^{+5e-9}_{-5e-9}$ & $13.989^{+5e-9}_{-5e-9}$  \\
		\midrule
  RG(Sly2) & 11.76 & $11.751^{+3e-7}_{-3e-7}$ & $11.751^{+8e-7}_{-8e-7}$ & 10.71 & $10.689^{+5e-7}_{-5e-7}$ & $10.689^{+1e-6}_{-3e-6}$  \\
		\midrule
  APR & 11.34 & $11.323^{+9e-5}_{-9e-5}$ & $11.323^{+0.001}_{-0.001}$ & 10.87 & $11.281^{+7e-5}_{-7e-5}$ & $11.281^{+1e-4}_{-1e-4}$  \\
		\midrule\bottomrule
	\end{tabular}
	\caption{Values obtained from the analysis for the NS radius of a \SI{1.4}{\Msun} and a \SI{2}{\Msun} star as shown in  \autoref{fig:NS_detector} with $N_{\rm ET} = 1000$ and the two different detector designs. The most probable values are also accompanied by the corresponding 1-$\sigma$ uncertainties. 
    The corresponding injected values are listed in columns 2, 5 for each of the EoS models. }
	\label{tab:NS_radius_eos_detector}
\end{table*}

\begin{table*}[tbh]
	\setlength\extrarowheight{4pt}
	\begin{tabular}{||c|c|c|c|c|c|c||}
		\toprule\midrule
		
		EoS &  \multicolumn{3}{c|}{$\Lambda_{1.4{\rm M}_{\odot}}$} &  \multicolumn{3}{c|}{$\Lambda_{2{\rm M}_{\odot}}$}\\ [0.5ex]
		\midrule
  & Injected & 2L \SI{15}{\kilo\meter}& $\Delta$ \SI{10}{\kilo\meter}& Injected & 2L \SI{15}{\kilo\meter}& $\Delta$ \SI{10}{\kilo\meter}\\
  \midrule
  PCP(BSk24) &  518.3 & $512.10^{+3.18}_{-0.37}$ & $512.65^{+2.63}_{-0.92}$ & 40.6 & $37.19^{+0.09}_{-0.25}$ & $37.26^{+0.01}_{-0.32}$\\
		\midrule
  GPPVA(NL3$\omega\rho$) & 936.7 & $931.54^{+2e-6}_{-2e-6}$ & $931.54^{+2e-6}_{-2e-6}$ & 118.8 & $114.96^{+2e-7}_{-2e-7}$ & $114.96^{+2e-7}_{-2e-7}$ \\
		\midrule
  RG(Sly2) &  309.0 & $306.15^{+3e-6}_{-3e-6}$ & $306.15^{+8e-6}_{-8e-6}$ & 11.4 & $11.16^{+8e-6}_{-8e-6}$ & $11.16^{+2e-5}_{-2e-5}$ \\
		\midrule
  APR &  248.0 & $266.28^{+0.01}_{-0.01}$ & $266.28^{+0.02}_{-0.02}$ & 14.7 & $22.38^{+0.003}_{-0.003}$ & $22.38^{+0.004}_{-0.004}$ \\
		\midrule\bottomrule
	\end{tabular}
	\caption{Same as \autoref{tab:NS_radius_eos_detector} for the NS tidal deformabilities.  }
	\label{tab:NS_lambda_eos_detector}
\end{table*}
\begin{figure*}[]
    \centering
    \includegraphics[width=1.\textwidth]{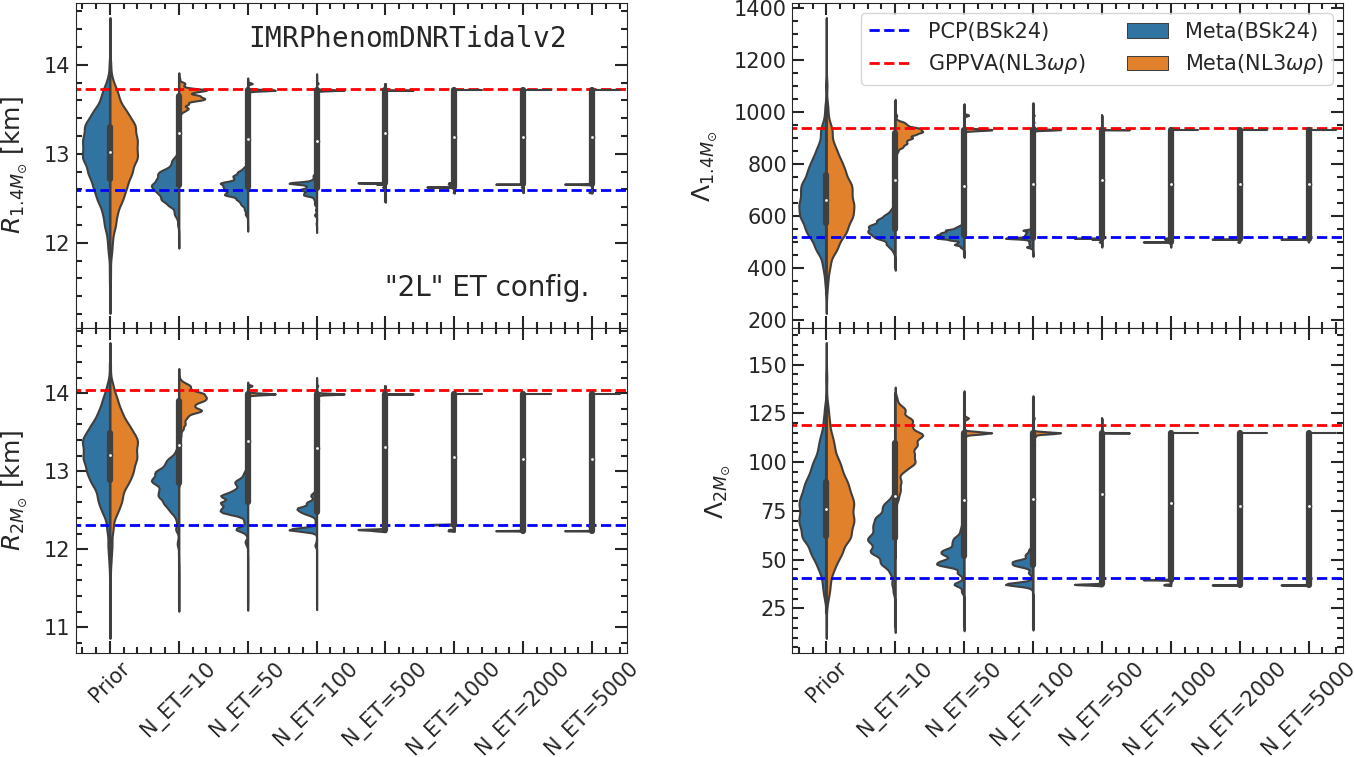}
    \caption{NS properties for PCP(BSk24) and GPPVA(Nl3$\omega\rho$) EoS obtained with different values of the number of detections, $N_{\rm ET}$, ranging from 10 to 5000.}
    \label{fig:NS_N_ET}
\end{figure*}
\begin{figure*}[tbh]
    \centering
    \includegraphics[width=1.\textwidth]
    {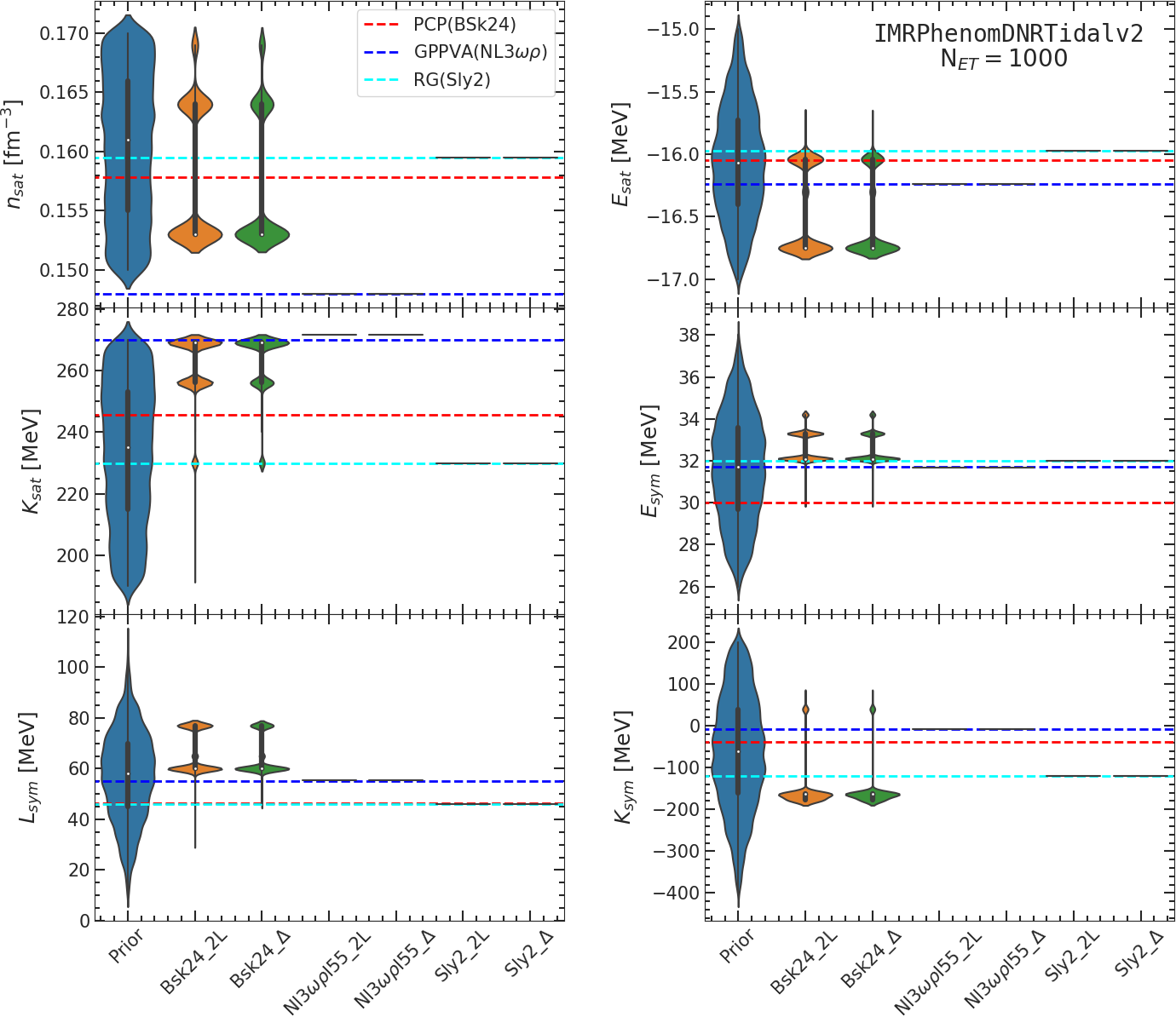}
    \caption{Nuclear matter parameters for detector configurations 2L and $\Delta$ obtained with $N_{\rm ET}=1000$.}
    \label{fig:NMP_detector}
\end{figure*}
\begin{figure*}[tbh]
    \centering
    \includegraphics[width=1.\textwidth]
    {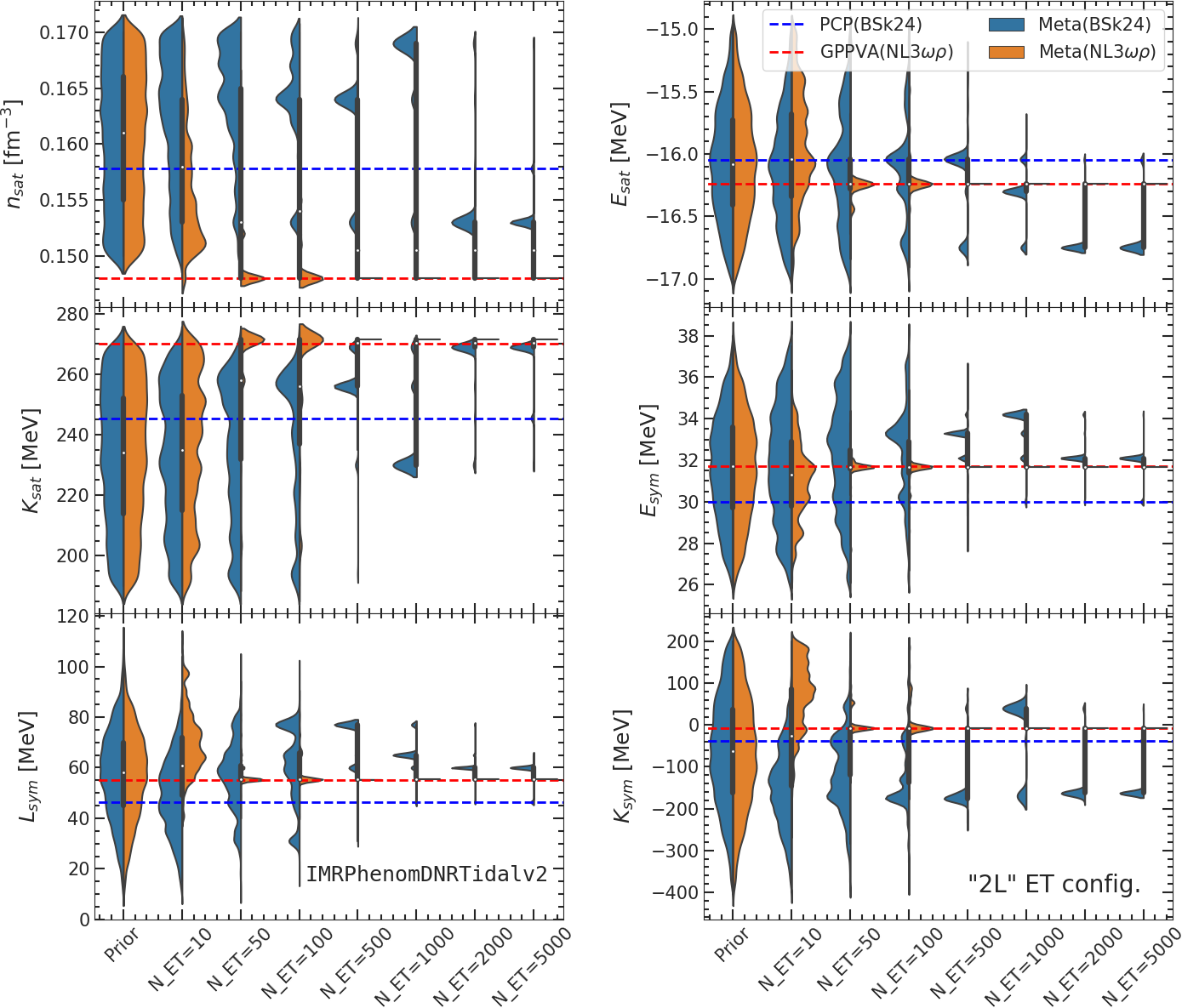}
    \caption{Nuclear matter parameters for PCP(BSk24) and GPPVA(NL3$\omega\rho$) EoS obtained with different $N_{\rm ET}$.}
    \label{fig:NMP_N_ET}
\end{figure*}

Even with 3G detectors, in our setup, it is extremely difficult to measure the individual tidal deformabilities of the two coalescing stars, except for very high--SNR events.\footnote{Taking into account more sophisticated treatments of the matter effects could improve on this, \textit{e.g.} in our case a difference in the errors among the waveforms including or not the first post--merger peak is present also on the individual tidal parameters, yet less pronounced than on $\tilde\Lambda$ (see \autoref{fig:AllEos_DelLam_pop_cumul}), in particular due to parameter degeneracy.}
A first question is then to which extent the individual tidal deformabilities
can be recovered from the different measured values of their combination $\tilde\Lambda$, defined in \autoref{eq:tildeLam_def}.
As representatives we choose the values of the tidal
deformability of a \SI{1.4}{\Msun} and a \SI{2}{\Msun} star,
$\Lambda_{1.4{\rm M}_{\odot}}$ and $\Lambda_{2{\rm M}_{\odot}}$, respectively. Since $\Lambda$
decreases with increasing mass, the absolute values are smaller for a
\SI{2}{\Msun} star and more difficult to determine. Thus, we expect that
the relative uncertainties will generally be larger for $\Lambda_{2{\rm M}_{\odot}}$ than for
$\Lambda_{1.4{\rm M}_{\odot}}$. An additional point is that our capability to determine
the tidal deformability of a star with a given mass will depend on the
mass distribution of the measured events, a dependence which is going to be
more pronounced if only a few events are detected. This can be seen from \autoref{fig:NS_iter}, where we show the results for the two fiducial values of the radius and the corresponding tidal deformabilities for a meagre number of ten detections, $N_{\rm ET} = 10$. The PCP(BSk24) EoS has been used for this study. The different violins correspond to different random choices for the ten detections out of $\ssim \num{6.9e4}$ 
simulated ones. The process was repeated over 50 iterations, out of which only a few representatives are displayed. In all cases we improve over our prior and the results remain compatible. The exact values and in particular the precision to which we are able to determine radii and individual tidal deformabilities depend, however,  on the exact detection sample, but this indeterminacy quickly vanishes as we increase the number of detections used in the posterior estimation. For the results shown, the 2L detector configuration has been chosen, but the conclusions are exactly the same with the triangle one. 

Let us now compare in more detail the detector configurations with a higher number of detections, $N_{\rm ET} = 1000$. 
In \autoref{fig:NS_detector} we show the same radii and tidal deformabilities, comparing the two detectors configurations and three different injected EoS models, PCP(BSk24), RG(SLy2) and GPPVA(NL3$\omega\rho$). 
The horizontal dashed lines indicate the respective injected values. On the scale of the figure, the injected values are very precisely reproduced and no difference is observed between the two detectors. Looking more closely at the corresponding numbers listed in \autoref{tab:NS_radius_eos_detector} and \ref{tab:NS_lambda_eos_detector}, we first confirm that there is no significant difference between the two detectors configurations in recovering the NS properties with the same number of detections, even though it is important to stress that the overall number of expected detections does depend on the different configurations, see \autoref{fig:AllEos_SNR_pop_cumul}.  
Second, the uncertainties are very small on the extracted values, but we do not exactly recover the injected ones, but there remains a tiny offset. This is the case for all the injected EoS and the two detector configurations. 
The reason is that the nuclear meta--model does not exactly contain the injected EoS, but only a representation which is very close and therefore we can only reproduce the injected values up to the precision to which the meta--model can represent the injected EoS. 
The APR EoS is somewhat an exceptional case. To start with, as shown in \autoref{fig:EoS_lamofm_plot}, its $\Lambda(m)$ relation does not lie inside the nuclear prior, in particular at high densities, \textit{i.e.} high NS masses, it predicts $\Lambda$ values lower than those covered by the nuclear prior. Similarly, the obtained radii for high masses are below the range of the nuclear prior. Further, the meta--model representation of the APR EoS becomes acausal much before reaching the \SI{2}{\Msun} NS. A dedicated study to understand this anomaly is in progress. It, however, explains the slight tension observed between the extracted and the injected values in particular for the \SI{2}{\Msun} case for the APR EoS. It is important, to note that RG(SLy2) is rather well recovered even though it has the same behaviour like APR in the $\Lambda$--$m$ plane.

Another point is that the uncertainties are much larger for the PCP(BSk24) EoS than for the others. This effect can be understood looking at the results for the PCP(BSk24) EoS as function of the number of detections, shown in \autoref{fig:NS_N_ET}. For a small number of detections, the uncertainties remain large and the extracted values are compatible with the injected ones. 
Increasing the number of detections, as expected, the extraction of radii and tidal deformabilities becomes more precise. There is, however, a multi--peaked structure formed with a competition between several most favored values, including the injected one. Still increasing the number of detections, the extracted results converge close to the injected ones. The reason for the multi--peaked structure is that within the nuclear meta--model there is a degeneracy between the nuclear matter parameters entering the model description and there are several combinations which allow equally well to recover the simulated detections for an intermediate number of detections. With a sufficiently high number of detections this structure disappears and the solution close to the injected one becomes favored. For the other EoS models, no such multi--peaked structure exists, see the GPPVA(NL3$\omega\rho$) one shown in \autoref{fig:NS_N_ET}. This can be explained by the fact that the PCP(BSk24) EoS lies in the middle of our nuclear prior whereas the other EoS chosen here for injection are at the upper or lower border of the nuclear prior, see \autoref{fig:EoS_lamofm_plot}. In this case the nuclear prior lifts the degeneracy between the meta--model parameters and enforces the solution close to the injected one. 
To make a closing argument, our results suggest that we will pin down the NS EoS with $\approx 500$ BNS events with sufficiently high SNR that lead to a tidal polarizability measurement. Considering the results of Fig.~\ref{fig:AllEos_SNR_pop_cumul}, this should be possible with all the proposed ET configurations.

\subsection{Reconstruction of the nuclear--physics parameters}\label{subsec:nuc_phys_params_results}

Concerning the extraction of information about nuclear matter, let us start by comparing the two detector configurations. In \autoref{fig:NMP_detector}, we display three isoscalar NMPs, the saturation density $n_{sat}$, the binding energy at saturation, $E_{sat}$ and the incompressibility $K_{sat}$ as well as three isovector ones, the symmetry energy $E_{sym}$, its slope $L_{sym}$ and the symmetry incompressibility $K_{sym}$. In addition to the nuclear prior, the extracted values for $N_{\rm ET} = 1000$ with injections from the PCP(BSk24), the GPPVA(NL3$\omega\rho$) and the RG(SLy2) EoS models are shown. The injected ones are indicated by the dashed horizontal lines. No significant difference between the two detector configurations can be observed. For GPPVA(NL3$\omega\rho$) and RG(SLy2) the injected values are perfectly recovered, whereas for PCP(BSk24), again a multi--peaked structure develops. This finding perfectly illustrates the degeneracy already discussed above: different combinations of nuclear parameters lead to the almost the same EoS for $\beta$--equilibrated NS matter and thus measuring the NS tidal deformability alone does not allow to determine the nuclear matter properties. Similar conclusions are discussed, \textit{e.g.}, in Ref.~\cite{Xie:2020tdo,Mondal:2021vzt} and in Ref.~\cite{Imam:2023ngm}, where a direct mapping of the nuclear matter parameters to a function $\Lambda(m)$ is attempted. This means that additional information ideally on symmetric matter is needed in order to pin down the nuclear matter properties, see the discussion in Ref.~\cite{Mondal:2021vzt}. For GPPVA(NL3$\omega\rho$) and RG(SLy2) this additional information comes from the nuclear prior in the sense that both being at the border of the prior distribution, their nuclear matter parameters are sufficiently constrained by the nuclear physics information entering that prior. Increasing the number of detections to up to $N_{\rm ET} = 5000$ does not considerably improve the situation, see the results for PCP(BSk24) in \autoref{fig:NMP_N_ET}, and additional information on symmetric matter is needed. On the other hand, the results for GPPVA(NL3$\omega\rho$) clearly show that, if we are able to additionally constrain the EoS for symmetric matter, the extraction of the nuclear matter parameters converges fast already for a moderate number of detections.\footnote{It should again be noted that part of the results presented in the current and previous section follow from our choice of a flat NS mass distribution: employing a Gaussian distribution for the masses as in \hyperref[footnote:massdistr_popres]{footnote~11} we find a considerable degradation in the accuracy in the extracted NS [cf. \autoref{fig:NS_detector}] and nuclear matter [cf. \autoref{fig:NMP_detector}] properties. This leads to the conclusion that information from more massive BNSs (such as GW190425~\cite{LIGOScientific:2020aai}) might turn out to be crucial.} This shows that a joint effort from the nuclear physics side ~\cite{HADES:2020lob,HADES:2022osk,Huth:2021bsp} and 3G GW detectors will allow us to determine the nuclear matter properties to extremely high precision within less than one year of operation for the latter. 

\section{Summary and conclusions}\label{sec:conclusion}
In this work we have performed a comparative study on two proposed designs of the Einstein Telescope, which is envisaged to be built within the next decade. We analyzed their potential impact on our understanding of the EoS of dense matter and the associated nuclear physics parameters. This is a further elaboration on the prospective nuclear physics goals to be addressed by  ET, out of many enthralling science cases, as outlined recently in Ref.~\cite{Branchesi:2023mws}. 

We simulated synthetic BNS coalescences with several state of the art EoS models from the \textsc{CompOSE} \cite{Typel:2013rza,CompOSECoreTeam:2022ddl} database along and up--to--date merger rate distribution. We obtained the statistical uncertainties on the observed sources' parameters employing the Fisher information matrix formalism though the \textsc{GWFAST} package  \cite{Iacovelli:2022mbg,Iacovelli:2022bbs}, and we tested the reliability of the Gaussian approximation with full parameter estimation studies on a few chosen events. An interesting point here is that the adherence of the FIM approach to a full PE for the tidal parameters depends on the chosen waveform approximant, with the phenomenological model considered showing the best agreement. An indication to explain this behaviour is given by how different models change when varying the tidal parameters, as discussed in \autoref{subsec:FIM_PE_comparison} and shown in \autoref{fig:mismatch_source1}; more in--depth checks, which go beyond the aim of the present work, would anyway be needed. Given the observed behaviour, we only relied on phenomenological models for our analyses. 

At the population level we have seen that the overall number of detections depend on the considered EoS model when adopting a flat mass distribution (with the EoS predicting the higher maximum masses resulting in the higher number of detections), with number of detections per year with ${\rm SNR}\geq12$ never being below $\order{\num{2.7e4}}\, {\rm events/yr}$, and we also find about $\ssim1.5$ more detections from the 2L design compared to the triangular one. Also, if post--merger effects of the first emission peak are incorporated in the waveform approximant, the number of detected events with smaller relative uncertainties in tidal deformabilities are found to be slightly higher (no significant decrease in the uncertainty of the very best events is anyway found). The signal--to--noise ratio and the associated number of detections, however, has no such dependence on the waveform model. \\
\par The extraction of neutron star properties, such as radius and tidal deformability at fiducial values of the mass, along with the underlying EoS used in the population generation was carried out with a nucleonic meta--modelling approach within a Bayesian framework. We observed that with $\gtrsim 500$ detections, one can recover with great accuracy the NS properties and the underlying injected EoS. 
These extractions are independent of the two considered geometries of the ET. Concerning the associated NMPs, we also demonstrated that they have some interdependent degeneracies, which could not be disentangled by using information only from $\beta$--equilibrated matter. Further experimental or observational information will be needed to improve this.  
Let us however stress that the achieved accuracy on the NS radii and the underlying EoS of $\beta$--equilibrated dense matter from the proposed ET would be unprecedented irrespective of the chosen geometry. 

\let\oldaddcontentsline\addcontentsline
\renewcommand{\addcontentsline}[3]{}
\acknowledgements
We thank Tanja Hinderer for useful comments on the manuscript. FG, CM and MO acknowledge financial support from the Agence Nationale de la recherche (ANR) under contract number ANR-22-CE31-0001-01. CM also acknowledges partial support from the Fonds de la Recherche Scientifique (FNRS, Belgium) and the Research Foundation Flanders (FWO, Belgium) under the EOS Project nr O022818F and O000422. The work of M.Maggiore and FI  is supported by the  Swiss National Science Foundation, grant 200020$\_$191957, and  by the SwissMap National Center for Competence in Research. FI is also supported by the Istituto Svizzero ``Milano Calling'' fellowship. AP is supported by the research programme of the Netherlands Organisation for Scientific Research (NWO). Part of his work was performed using the Computing Infrastructure of Nikhef, which is part of the research program of the Foundation for Nederlandse Wetenschappelijk Onderzoek Instituten (NWO-I), which is part of the Dutch Research Council (NWO). Part of the computations made use of the Baobab and Yggdrasil clusters at the University of Geneva.
The work of M.Mancarella is supported by European Union's H2020 ERC Starting Grant No.~945155--GWmining, Cariplo Foundation Grant No.~2021-0555, the ICSC National Research Centre funded by NextGenerationEU, and MIUR PRIN Grant No. 2022-Z9X4XS. TD acknowledges funding from the EU Horizon under ERC Starting Grant, no.\ SMArt-101076369, and support from the Deutsche Forschungsgemeinschaft, DFG, project number DI 2553/7.

\let\oldaddcontentsline\addcontentsline
\renewcommand{\addcontentsline}[3]{}
\bibliography{myrefs}

\end{document}